\documentclass[apj]{emulateapj}

\newcommand{\etal}{{et al.~}}
\newcommand{\msunh}{\>h^{-1}\rm M_\odot}
\newcommand{\Msun}{\>{\rm M_\odot}}

\newcommand{\mpch}{\>h^{-1}{\rm {Mpc}}}
\newcommand{\kms}{\>{\rm km}\,{\rm s}^{-1}}
\newcommand{\calC}{{\cal C}}
\newcommand{\rmag}{\>^{0.1}{\rm M}_r-5\log h}

\def\gtsima{$\; \buildrel > \over \sim \;$}
\def\ltsima{$\; \buildrel < \over \sim \;$}
\def\prosima{$\; \buildrel \propto \over \sim \;$}
\def\gsim{\lower.7ex\hbox{\gtsima}}
\def\lsim{\lower.7ex\hbox{\ltsima}}
\def\simgt{\lower.7ex\hbox{\gtsima}}
\def\simlt{\lower.7ex\hbox{\ltsima}}
\def\simpr{\lower.7ex\hbox{\prosima}}
\def\la{\lsim}
\def\ga{\gsim}
\def\lta{\la}
\def\gta{\ga}

\begin{document}

\shorttitle{The clustering of groups}
\shortauthors{Wang et al.}

\title{The clustering of SDSS galaxy groups: mass and color dependence}

\author{Yu Wang\altaffilmark{1,2,6}, Xiaohu Yang\altaffilmark{2,6}, H.J.
  Mo\altaffilmark{3}, Frank C. van den Bosch\altaffilmark{4}, Simone M.
  Weinmann\altaffilmark{5}, Yaoquan Chu\altaffilmark{1}}

\altaffiltext{1}{Center for Astrophysics, University of Science and
  Technology of China, 230026, P. R. China}
\altaffiltext{2}{Shanghai Astronomical Observatory, the Partner Group
  of MPA, Nandan Road 80, Shanghai 200030, China}
\altaffiltext{3}{Department of Astronomy, University of Massachusetts,
  Amherst MA 01003-9305}
\altaffiltext{4}{Max-Planck Institute for Astronomy, D-69117 Heidelberg,
  Germany}
\altaffiltext{5}{Max-Planck-Institut f\"ur Astrophysik,
  Karl-Schwarzschild-Strasse 1, 85748 Garching, Germany}
\altaffiltext{6}{Joint Institute for Galaxy and Cosmology
  (JOINGC) of Shanghai Astronomical Observatory and University of
  Science and Technology of China}

\begin{abstract} We use a large sample of galaxy groups selected from the SDSS
  data release  4 with an  adaptive halo-based group  finder to probe  how the
  clustering  strength of  groups  depends  on their  masses  and colors.   In
  particular, we determine the relative  biases of groups of different masses,
  as  well as that  of groups  with the  same mass  but with  different colors
  (either  that  of the  central  galaxy,  or the  total  color  of all  group
  members).  In  agreement with  previous studies, we  find that  more massive
  groups are more strongly clustered,  and the inferred mass dependence of the
  halo  bias  is in  good  agreement  with  predictions for  the  $\Lambda$CDM
  concordance cosmology.  Regarding the  color dependence, we find that groups
  with red centrals  are more strongly clustered than groups  of the same mass
  but with  blue centrals.  Similar results  are obtained when the  color of a
  group is  defined to be the total  color of its member  galaxies.  The color
  dependence   is  more  prominent   in  less   massive  groups   and  becomes
  insignificant in groups with masses $\gta 10^{14}\msunh$.  These results are
  consistent with those obtained by Yang \etal from an analysis of the 2dFGRS,
  but inconsistent with those obtained by Berlind {\etal},
  who also used an SDSS
  group  catalogue.  We construct  a mock  galaxy redshift  survey
  from the large Millennium $N$-body simulation that is populated with galaxies
  according to  the semi-analytical model  of Croton \etal Applying  our group
  finder to this mock survey, and  analyzing the mock data in exactly the same
  way as the  true data, we are able to accurately  recover the intrinsic mass
  and color  dependencies of the halo  bias in the  model.  Interestingly, the
  semi-analytical model reveals the same  color dependence of the halo bias as
  we  find in  our  group  catalogue.  In  halos  with $M\sim  10^{12}\msunh$,
  though, the strength  of the color dependence is much  stronger in the model
  than  in  the  data.
  We discuss these
  results in  light of  the assembly bias  of dark  matter halos and  the star
  formation histories of galaxies.
\end{abstract}



\section{introduction}
\label{sec:intro}

In  the  standard  cold   dark  matter  (CDM)  paradigm  of  structure
formation,  virialized CDM  halos are  considered to  be  the building
blocks of  the mass  distribution in the  Universe. The  properties of
dark matter halos, as well as their formation histories and clustering
properties,  have been studied  in great  detail using  both numerical
simulations  and analytical  approaches such  as the  (extended) Press
Schechter (1974) formalism. These studies have shown that halo bias is
mass dependent, in that more massive halos are more strongly clustered
(e.g., Mo  \& White 1996;  Sheth \& Tormen  1999; Sheth, Mo  \& Tormen
2001; Seljak \& Warren 2004;  Tinker \etal 2005). This mass dependence
of the halo bias has played a crucial role in our understanding of the
correlation  function  of  both  dark  matter and  galaxies,  via  the
halo model (e.g., Cooray \& Sheth 2002), the halo occupation
model (e.g., Jing, Mo \&  B\"orner 1998; Seljak 2000; Peacock \& Smith
2000;  Berlind  \&  Weinberg  2002; Magliocchetti  \&  Porciani  2003;
Berlind \etal 2003; Zheng  \etal 2005), and the conditional luminosity
function (CLF)  model (e.g., Yang, Mo  \& van den Bosch  2003; van den
Bosch, Yang  \& Mo 2003; Vale  \& Ostriker 2006; Cooray  2006; van den
Bosch \etal 2007).

Observationally, the clustering strength of galaxy systems (groups and
clusters) is  found to increase  with their mean separation  $d \equiv
n^{-1/3}$, with  $n$ the number  density of objects (e.g.   Bahcall \&
West  1992; Zandivarez  \etal  2003; Padilla  \etal  2004; Yang  \etal
2005b). Assuming that richer  groups are statistically associated with
more massive halos, one can  estimate the halo masses of galaxy groups
using a theoretical  halo mass function, such as the  one given by the
current $\Lambda$CDM model (see Yang \etal 2005b; Berlind \etal 2006a;
Yang  \etal 2007).   Thus  the relation  between  mean separation  and
clustering strength can be converted into a relation between halo mass
and halo bias.  As shown in Yang \etal (2005b), the mass dependence of
the  halo  bias  obtained this  way  from  galaxy  groups is  in  good
agreement with the prediction of the current $\Lambda$CDM model.

In addition to the mass dependence, recent simulations have shown that
halo  bias  also depends  on  the  assembly  history.  Using  $N$-body
simulations,  Sheth   \&  Tormen  (2004)   studied  the  environmental
dependence of halo assembly and  found that dark matter halos in dense
environments assemble at slightly earlier times than halos of the same
mass  in low density  regions.  Gao  \etal (2005)  used a  very large,
high-resolution  numerical  simulation  of  structure formation  in  a
$\Lambda$CDM  cosmology to reexamine  the mass  and age  dependence of
halo bias.  They  found that, for halos of $M  \leq 10^{13} \msunh$ at
redshift $z=0$, the bias factor depends not only on halo mass but also
on the halo assembly time. This  result has been confirmed by a number
of subsequent investigations (Harker  \etal 2006; Wechsler \etal 2006;
Wetzel \etal  2006; Gao \&  White 2007; Jing,  Suto \& Mo  2007).  The
origin of this assembly bias has recently been investigated
in a number of papers (Wang, Mo \& Jing 2007; Keselman \& Nusser 2007;
Hahn \etal 2007).

If the  properties of galaxies that reside  in haloes of  a given mass
correlate with  the assembly history  of that halo, the  assembly bias
will impact the galaxy bias, $b_{\rm gal}$, to the extent that $b_{\rm
  gal}$ not only depends on halo mass, as generally assumed in HOD and
CLF modeling, but also on the formation history of that halo.  Using a
very crude method  for  assigning galaxy  luminosities to dark  matter
haloes,    Reed    \etal  (2007)    conclude,    however,   that   the
luminosity-weighted  galaxy ages  cannot   closely trace the  assembly
epoch of  their    dark   matter  hosts,  otherwise     the   observed
color-dependence of  the clustering  can  not be reproduced.  Using  a
galaxy formation model grafted    on  to the   Millennium  Simulation,
Croton, Gao  \&  White  (2007) compared   the original   simulation to
`shuffled' versions  in   which the galaxy  populations   are randomly
swapped among  haloes of similar  mass. They  found that  the shuffled
versions have   correlation functions  that  are  different  from  the
original ones at  the 5 to   10 percent level, indicating  that galaxy
properties are (weakly)   correlated with the halo assembly   history.
Unfortunately, since their    semi-analytical model is not a   perfect
representation of the true  universe  (see e.g., Weinmann \etal  2006b
and Baldry \etal 2006), it is  unclear whether such a correlation also
exists in reality.

To test this, various authors  have examined whether there is any sign
of assembly bias  in the large scale structure  of the observed galaxy
distribution.   Using a  galaxy group  catalogue constructed  from the
2-degree Field  Galaxy Redshift  Survey (2dFGRS; Colless  \etal 2001),
Yang  \etal (2006)  examined  how the  clustering  strength of  galaxy
groups depends on the star formation activity of the central galaxies.
At fixed mass  they found that the bias of  galaxy groups decreases as
the SFR of the central galaxy increases. Assuming that galaxies with a
more active SFR are bluer, this implies that groups with a red central
are more  strongly clustered than groups  of the same mass  but with a
blue central.  Surprisingly, a  completely opposite trend was found by
Berlind  \etal   (2006a) using  the  group  catalogue constructed  by
Berlind \etal (2006b) from the Sloan Digital Sky Survey (SDSS; York et
al.  2000).   They found that  massive groups with bluer  centrals are
{\it more}  strongly biased  on large scales  than groups of  the same
mass  but  with  redder   centrals.   Tinker  \etal  (2007)  used  the
(projected) galaxy correlation function  and galaxy void statistics to
test  whether   the  galaxy  content   of  halos  of  fixed   mass  is
systematically different in low density environments.  Using data from
the SDSS combined  with HOD models, they find  that the luminosity and
color of  field galaxies are  determined predominantly by the  mass of
the halo in which they reside and have little direct dependence on the
environment in which the host halo formed. Since, as shown by Sheth \&
Tormen (2004), the  assembly history of a halo  is correlated with its
environment,  this  therefore implies  that  there  is no  significant
correlation between the color of  a galaxy and the assembly history of
its dark matter halo.   Finally, Blanton \& Berlind (2006)compared the
actual  relationship between  galaxy  colors and  large scale  density
field to that predicted by  the null hypothesis that galaxy colors are
only dependent on the mass of  the halo in which they reside.  To this
extent, they use a similar shuffling technique as Croton \etal (2007),
but using galaxy groups in the SDSS rather than dark matter halos in a
semi-analytical model  for galaxy formation.  They  find that shuffled
and unshuffled color-density relations agree to better than 5 percent,
indicating that the large scale environment of a group has only a very
mild impact  on the properties  of its member galaxies.   Clearly, the
current status as to whether galaxy properties are correlated with the
halo assembly history is still very confusing.

In this paper, we re-examine the  dependence of the halo bias on group
mass  and on the  colors of  member galaxies,  using the  galaxy group
catalogue recently constructed by Yang  \etal (2007) from the SDSS DR4
(Adelman-McCarthy  \etal  2006).   An  important improvement  of  this
catalogue over that of Yang \etal (2006) and Weinmann \etal (2006a) is
that group masses are estimated both from the total luminosity and the
total stellar mass of member  galaxies, allowing us to examine how the
results may be affected by  the uncertainties in the estimate of group
masses.  Furthermore, in order to test the impact of the uncertainties
in the  group finder and in the  methods used to assign  masses to our
groups,  we  use  a  mock  galaxy  redshift  survey  (hereafter  MGRS)
constructed   from  the   Millennium  semi-analytical   galaxy  sample
(hereafter SAM) of  Croton \etal (2006) to test  how accurately we can
recover the true trends in the semi-analytical model.

This paper is  organized as follows.  Section~\ref{sec:data} describes
the galaxy  sample and  group catalogue used  for our  analysis, while
Section~\ref{sec:method}  presents  our   method  for  estimating  the
group-galaxy  cross-correlation function.   Our  observational results
for the dependence of the group-galaxy cross-correlation on group mass
and  on the  color  of the  central  galaxy are  presented in  Section
\ref{sec:obs}.  The same results. but for the semi-analytical MGRS are
presented   in   Section    \ref{sec:MGRS}.    Finally,   in   Section
\ref{sec:summary}, we summarize  and discuss our findings.  Throughout
this  paper, we  adopt the  $\Lambda$CDM `concordance'  cosmology with
$\Omega_m    =    0.24$,    $\Omega_{\Lambda}=0.76$,   $h=0.73$    and
$\sigma_8=0.77$.  Distances are quoted in units of $\mpch$.

\begin{deluxetable*}{lcccc}
  \tablecaption{Volume-limited Samples of Galaxies\label{tab1}}
  \tablewidth{0pt} \tablehead{Sample & z &$\rmag$ & $N_{galaxy}$
    (obs.) & $N_{galaxy}$ (mock) }

  \startdata
  V1 & (0.064, 0.127) & (-22.0, -20.5] &  54116 & 66484 \\
  V2 & (0.064, 0.157) & (-22.0, -21.0] &  39101 & 51627
  \enddata

  \tablecomments{Column 1  indicates the volume-limited  galaxy sample
    ID.  Column 2  gives the redshift range of  each sample.  Column 3
    is the absolute-magnitude range.  Columns  4 and 5 list the galaxy
    number in each sample  for observational and mock SDSS catalogues,
    respectively.}
\end{deluxetable*}

\begin{deluxetable*}{lccccccccc}
\tabletypesize{\scriptsize}
  \tablecaption{Galaxy Groups in the Observation \label{tab2}}
  \tablewidth{0pt} \tablehead{ID & $\log M_{\rm h}$ & z &
    $N_{group}$ &\multicolumn{3}{c}{C0 ($M_L$/$M_S$)}
    &&\multicolumn{2}{c}{C1 ($M_L$/$M_S$)} \\
    \cline{5-7}\cline{9-10}& & & ($M_L$/$M_S$) & Red &
    Green & Blue && Red & Blue \\
    (1)&(2)&(3)&(4)&(5)&(6)&(7)&&(8)&(9)}

  \startdata
G1& $(12.0,12.5]$&(0.064,0.127)& 43233/41314& 0.99/0.99 &0.87/0.88 &0.60/0.66 &&0.94/0.96 &0.65/0.77\\
G2& $(12.5,13.0]$&(0.064,0.127)& 15614/15450& 1.01/1.01 &0.91/0.92 &0.67/0.72 &&0.96/0.99 &0.73/0.84\\
G3& $(13.0,13.5]$&(0.064,0.157)& 9909/10104& 1.03/1.03 &0.94/0.95 &0.71/0.75 &&0.99/1.01 &0.80/0.88 \\
G4& $(13.5,14.0]$&(0.064,0.157)& 2976/2826& 1.04/1.04 &0.96/0.97 &0.73/0.75 &&0.99/1.01  &0.86/0.89
  \enddata

  \tablecomments{Column  1 indicates  the  group sample  ID. Column  2
    indicates the  mass range of each  group sample in  terms of $\log
    [M_h/\msunh]$.  Column 3 gives the  redshift range of each sample.
    Column 4 lists the group  number in the corresponding redshift and
    mass  ranges, where $M_L$  and $M_S$  mean group  masses estimated
    from the  ranking of group luminosity $L_{19.5}$  and stellar mass
    $M_{\rm  stellar}$, respectively (Yang  \etal 2007).   Columns 5-7
    list the means  of the central galaxy colors  (C0) for groups with
    red, green  (galaxies between red  and blue Gaussian  peaks), blue
    central galaxies, respectively.  Columns 8 and 9 list the means of
    the total  group colors (C1)  for groups which are  separated into
    red and blue subsamples in similar numbers.  See the text for more
    details. }
\end{deluxetable*}

\begin{deluxetable*}{lccccccccc}
\tabletypesize{\scriptsize}
  \tablecaption{Galaxy Groups in the MGRS \label{tab3}} \tablewidth{0pt}
  \tablehead{ID & $\log M_{\rm h}$ & z & $N_{group}$ &\multicolumn{3}{c}{C0
      ($M_L$/$M_S$)} &&\multicolumn{2}{c}{C1 ($M_L$/$M_S$)}\\
    \cline{5-7}\cline{9-10}& & & ($M_L$/$M_S$) & Red & Green & Blue && Red & Blue\\
    (1)&(2)&(3)&(4)&(5)&(6)&(7)&&(8)&(9)}
\startdata
  S1& $(12.0,12.5]$& (0.064,0.127) & 46274/44761& 0.98/0.99 &0.86/0.86 &0.53/0.58 &&0.89/0.90 &0.54/0.61 \\
  S2& $(12.5,13.0]$& (0.064,0.127) & 16929/16538& 0.98/0.98 &0.87/0.89 &0.57/0.61 &&0.90/0.93 &0.59/0.68 \\
  S3& $(13.0,13.5]$& (0.064,0.157) & 11139/11231& 0.97/0.97 &0.89/0.91 &0.55/0.60 &&0.92/0.94 &0.61/0.69 \\
  S4& $(13.5,14.0]$& (0.064,0.157) & 3311/3453& 0.97/0.97 &0.92/0.93 &0.47/0.63 &&0.93/0.94 &0.53/0.68
\enddata

\tablecomments{Similar to  Table~\ref{tab2}, but for  galaxy groups in
  the semi-analytical Mock Galaxy Redshift Survey (MGRS).}
\end{deluxetable*}

\begin{figure*}
  \plotone{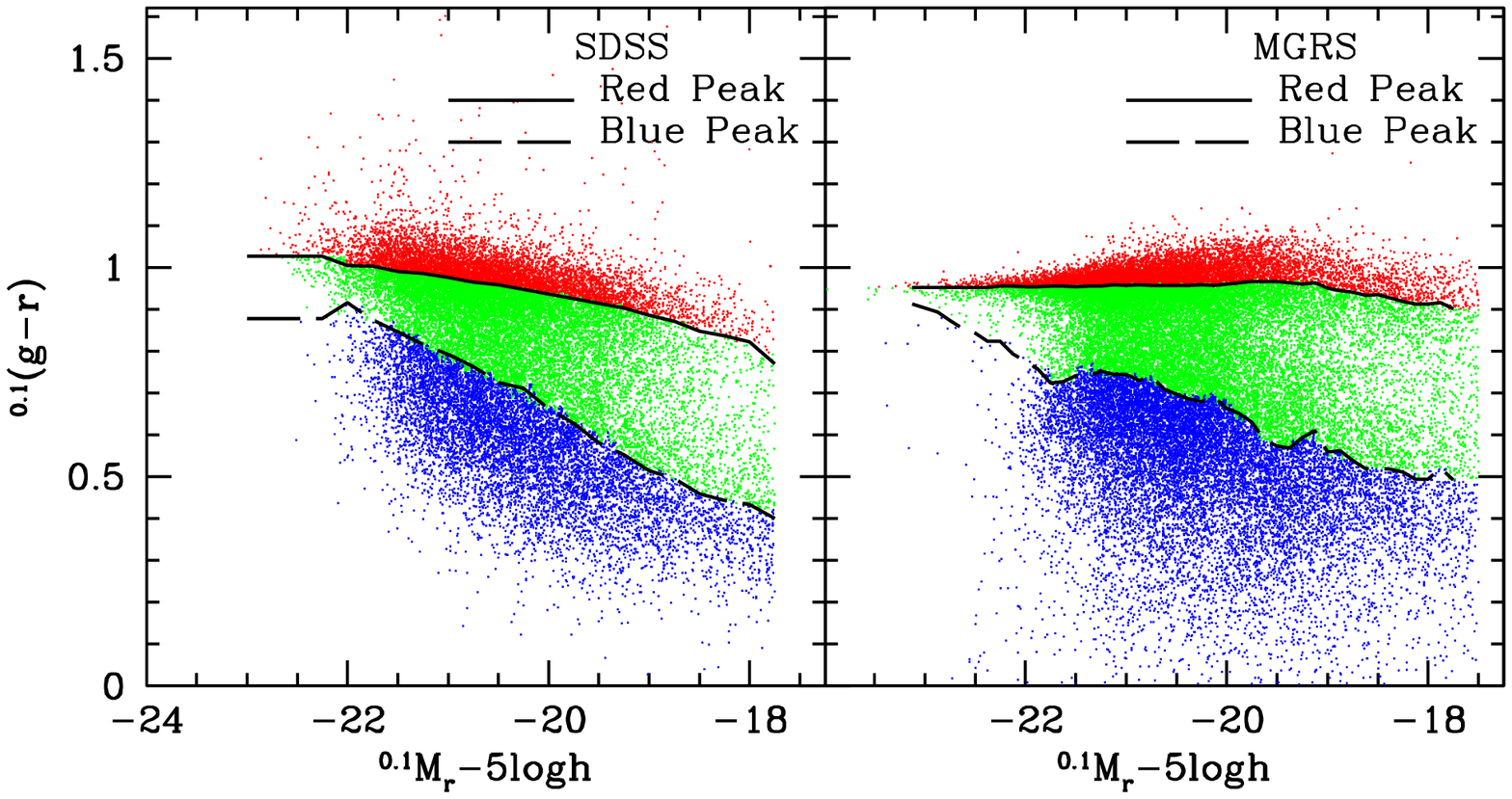}   \caption{The  color-magnitude   distributions  of
    galaxies in  the SDSS (left-hand  panel) and the  MGRS (right-hand
    panel). The dotted- and dashed-  lines indicate the centers of the
    two Gaussian used  to describe the color distributions  at a given
    luminosity bin (see text for details). These are used to split the
    galaxy populations  in red (above  the red sequence),  blue (below
    the  blue  sequence)  and  green  (in between  the  red  and  blue
    sequences). For clarity,  only 10\% of all
    galaxies in the SDSS and MGRS are shown. }
  \label{col:gau}
\end{figure*}

\section{Galaxy and Group Samples Used}
\label{sec:data}

\subsection{The Observational Samples of Galaxies}

We  use  galaxy  samples  constructed  from the  New  York  University
Value-Added Galaxy Catalogue (NYU-VAGC, see Blanton \etal 2005b).  The
version of the catalogue used here is based on the SDSS Data Release 4
(Adelman-McCarthy \etal 2006).
Only    galaxies    with   redshifts    $0.01\leq    z   \leq    0.2$,
extinction-corrected  Petrosian  magnitudes  $14.5  <  r  <17.6$,  and
redshift  completeness $\calC  >0.7$ are  selected.  The  $r$-band and
$g$-band  absolute  magnitudes  are  corrected to  $z=0.1$  using  the
$K$-correction    code   of    Blanton   \etal    (2003a)    and   the
luminosity-evolution   model  of  Blanton   \etal  (2003b).    In  the
apparent-magnitude limits  adopted here, the  bright end is  chosen to
avoid  the incompleteness  due to  the large  angular sizes  of nearby
galaxies, while the  faint end is chosen to  match the magnitude limit
of the SDSS main galaxy sample.  In addition, we also include galaxies
with $0.01\leq  z \leq 0.2$ in  the NYU-VAGC that  have redshifts from
alternative sources:  2dFGRS (Colless \etal 2001),  the PSCz (Saunders
\etal  2000) or  RC3 (de  Vaucouleurs \etal  1991).  Our  final sample
consists  of  286563 galaxies.   In  our  analysis,  we construct  two
volume-limited samples in two  bins of the absolute magnitude $\rmag$:
V1 (-22.0, -20.5] and V2  (-22.0, -21.0] (see Table~\ref{tab1} for the
detailed selection criteria). Note that these two volume-limited
galaxy samples are different from those used in constructing the
group catalogues, and are used only in measuring the cross correlations.

\subsection {Mock Samples of Galaxies}

In order to  check how accurately our methodology  allows us to detect
true effects  of assembly  bias, we construct  a mock  galaxy redshift
survey  (MGRS)  that  incorporates  the same  observational  selection
effects  as the  SDSS  sample.  Our  MGRS  is based  the model  galaxy
catalogue  constructed by  Croton \etal  (2006) using  a semi-analytic
model  of galaxy formation  in combination  with the  `Millennium Run'
$N$-body     simulation    (Springel     \etal     2005)    \footnote{
  http://www.mpa-garching.mpg.de/galform/agnpaper}.   The cosmological
parameters  adopted  in  the  `Millennium  Run'  are  $\Omega_m=0.25$,
$\Omega_\Lambda=0.75$, and a CDM  spectrum with an amplitude specified
by $\sigma_8=0.9$.  The simulation was performed with the code GADGET2
(Springel  \etal 2005),  using  $2160^3$ dark  matter  particles in  a
periodic  cubic box  with  a side  length  $L_{\rm box}=500\mpch$  (in
comoving  units).  The mass  of a  particle is  $8.6\times10^8\msunh$.
The galaxy catalogue is generated  based on a semi-analytical model of
galaxy  formation which  uses the  simulated halo  merging  trees.  We
refer  the  reader to  Croton  \etal (2006)  for  the  details of  the
semi-analytic  model.  Because of  the finite  mass resolution  in the
simulation,   the   sample  is   complete   to   a  luminosity   limit
$\rmag\sim-16.6$.

Our construction of  the MGRS here is similar to  that described in Li
\etal (2007a;  see also  Yang \etal 2004).   First, we  stack $3\times
3\times  3$ replicates  of  the  simulation box  and  place a  virtual
observer at  the center  of the stacked  boxes.  Next, we  assign each
galaxy a ($\alpha$, $\delta$)-coordinate  and remove the ones that are
outside the mocked  SDSS survey region.  For each  model galaxy in the
survey  region, we  compute its  redshift (which  include  the general
expansion,   the   peculiar   velocity,   and  a   $35\kms$   Gaussian
line-of-sight velocity dispersion to  mimic the redshift errors in the
data),  its  $r$-band  apparent   magnitude  (based  on  the  $r$-band
luminosity of the galaxy), and its absolute magnitude $\rmag$ which is
$K+E$ corrected  to $z=0.1$.  We  eliminate galaxies that  are fainter
than   the  SDSS   apparent-magnitude  limit,   and   incorporate  the
position-dependent  incompleteness  by  randomly eliminating  galaxies
according to  the completeness factors obtained from  the survey masks
that are provided  by the NYU-VAGC (Blanton \etal  2005b).  Finally we
select the same  two volume-limited samples (V1 and  V2) from the MGRS
as we  used for the  actual data.  The  numbers of (mock)  galaxies in
these two samples  are listed in the last column  of Table \ref{tab1}.
Note that they  do not match the SDSS data  perfectly. This mainly owes
to the fact that the semi-analytical model does not match the observed
luminosity function  perfectly.

\subsection {Random Samples of Galaxies}

In order to measure the  two-point correlation functions, one needs to
construct random samples to normalize the pair counts.  To do this, we
first generate a set of  random `galaxies' according to the luminosity
function  obtained by  Blanton \etal  (2003b).   Here a  large set  of
points  are randomly distributed  within the  survey sky  coverage and
each point is assigned a  redshift and an absolute magnitude according
to the luminosity  function. The redshifts are assigned to random
galaxies by assuming that they have a constant number density as a
function of redshift. Similar to the last step in
constructing the  MGRS, we apply the completeness  and magnitude limit
according to the  survey masks provided by Blanton  \etal (2005b). The
total number of `galaxies' in each random sample is about 7.5 times as
large as that in the corresponding observational or mock sample.

\subsection{Galaxy Group Samples}
\label{sec_groupsamples}

Our analysis  is based  on the group  catalogue of Yang  \etal (2007),
which is  constructed from  the SDSS Data  Release 4 using  a modified
version of  an adaptive  halo-based group finder  (Yang \etal  2005a).
This  group  finder  is  optimized  to  assign  galaxies  into  groups
according to  their common dark  matter halos. Each group  is assigned
two  values  of halo  mass,  one  is based  on  the  ranking of  group
characteristic luminosity  and is referred to  as the luminosity-based
halo mass ($M_L$), and the other is based on the total stellar mass of
the  group and  is referred  to as  the stellar  mass-based  halo mass
($M_S$)  (see Yang  \etal 2007  for details).   Note that  this method
requires  knowledge  of  the  halo  mass function,  and  is  therefore
cosmology dependent.

In this paper, we  measure the galaxy-group cross-correlation function
(hereafter  GGCCF) to  quantify the  dependence of  the  clustering on
group mass  and on  the color of  the group  members. To this  end, we
divide galaxy groups into four samples according to their halo masses.
The   criteria  used   to   define  these   samples   are  listed   in
Table~\ref{tab2}.  We  further divide each  of the group  samples into
color subsamples using either the color of the central galaxy (C0) or
using the  total color  of all group  members (C1). Below  we describe
these color subsamples in more detail.

\begin{itemize}
\item {\bf C0  subsamples:} based on the color of  the central galaxy.
  Here, the central  galaxy of a group is defined  to be the brightest
  member in the group, and we  assume that the location of the central
  galaxy coincides with  the center of mass of  the group.  Tests show
  that our results  do not change significantly if  the central galaxy
  is  defined to be the  most massive  group member instead. Note that
  only  a very small portion ($\la 0.6\%$) of groups in which the most
  massive  galaxy is different from the brightest galaxy.   Recent
  investigations of the  color-magnitude distribution of galaxies show
  that the colors of galaxies for a given absolute magnitude follows a
  bi-normal  form (e.g., Baldry  \etal 2004;  Blanton \etal  2005a; Li
  \etal  2006).  Following  Li  \etal  (2006), we  separate  the  SDSS
  galaxies  into  118 bins  according  to  their absolute  magnitudes,
  $\rmag$, and  fit the $^{0.1}(g-r)$  color distribution in  each bin
  using a double-Gaussian function.   The centers of the Gaussians are
  shown  in the  left-hand panel  of Fig.~\ref{col:gau}  as  solid and
  dashed lines,  and are used the  split the galaxy  population in red
  (above solid  line) blue (below  dashed line) and green  (in between
  the  solid and  dashed lines)  subsamples.  Using  the color  of the
  central galaxies, we also split the group catalogue in red, blue and
  green subsamples.  The  mean colors of the central  galaxies in each
  subsample are listed in columns 4 - 6 of Table ~\ref{tab2}.
\item {\bf C1 subsamples:} based on the total color of each group.  In
  this case, we first add up the $r$-band and $g$-band luminosities of
  all member galaxies for each group to get the total group luminosity
  $L_{r,t}$, $L_{g,t}$,  respectively.  The groups  are then separated
  into  red and  blue  subsamples  of equal  size  (i.e., with  equal
  numbers of  groups) according to the group  color $2.5*(\log L_{r,t}
  -\log L_{g,t})$.  The mean colors  of the groups in  each subsample
  defined  in  this  way are  listed  in  columns  7  and 8  of  Table
  ~\ref{tab2}.
\end{itemize}

We have  also selected galaxy groups  from the MGRS  using exactly the
same group-finding algorithm  as applied to the real  sample (see Yang
et al.  2007 for more details).  Here again we assign to each selected
group  two values  of mass,  one ($M_L$)  based on  the characteristic
group  luminosity and  the other  ($M_S$) based  on the  group stellar
mass.  The groups selected from the MGRS are also divided into samples
of different masses and into subsamples according to the color of the
central galaxy and  according to the total group  color, respectively.
The   information  about  about   these  samples   can  be   found  in
Table~\ref{tab3} with  the same  format as in  Table~\ref{tab2}.  When
modeling the  color-magnitude distribution  of the mock  galaxies, we
separate  the  galaxies   into  113  bins  of  $\rmag$   and  fit  the
$^{0.1}(g-r)$ color  distribution in  each bin with  a double-Gaussian
function.   The right-hand panel  of Fig  \ref{col:gau} shows  the two
ridges defined by  the peaks of the double-Gaussian  function.  As for
the SDSS data,  these ridges again are used  to separate galaxies into
red, green and blue subsamples.

\section{The Group-Galaxy Cross-Correlation Function}
\label{sec:method}

\begin{figure*}
  \plotone{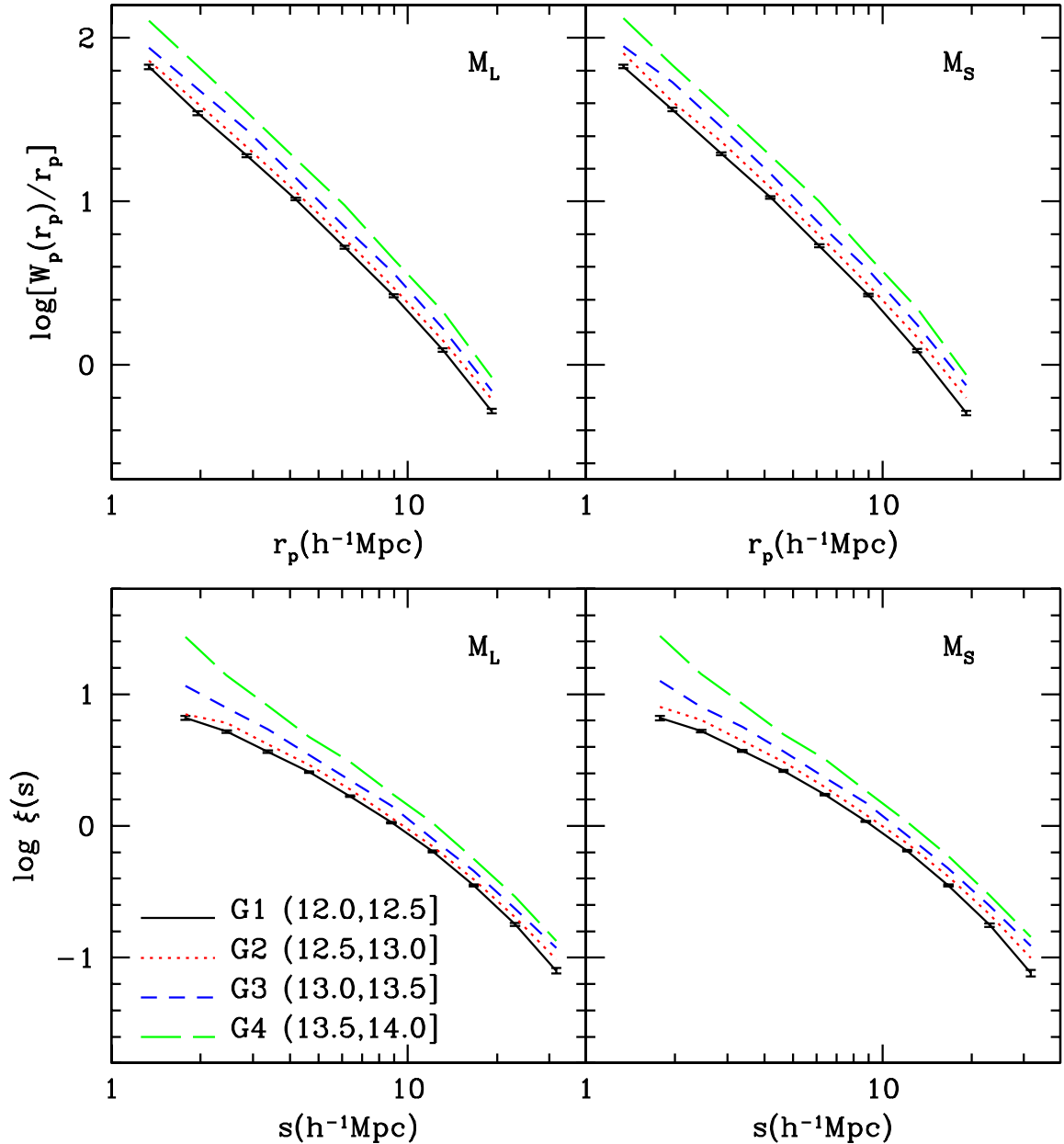} \caption{ Upper panels: the projected GGCCF between
    the volume-limited galaxy  sample V2 (-22.0,  -21.0] and the group
    samples (G1-G4) within different  mass ranges as indicated.  Lower
    panels: similar  to the upper   panels, but for  GGCCF measured in
    redshift space.  In the plot, $M_L$ (left panels) and $M_S$ (right
    panels) correspond to groups  with masses that are estimated  from
    the ranking of group luminosity and stellar mass, respectively. }
\label{fig:xi_obs}
\end{figure*}

\begin{figure*}  \plotone{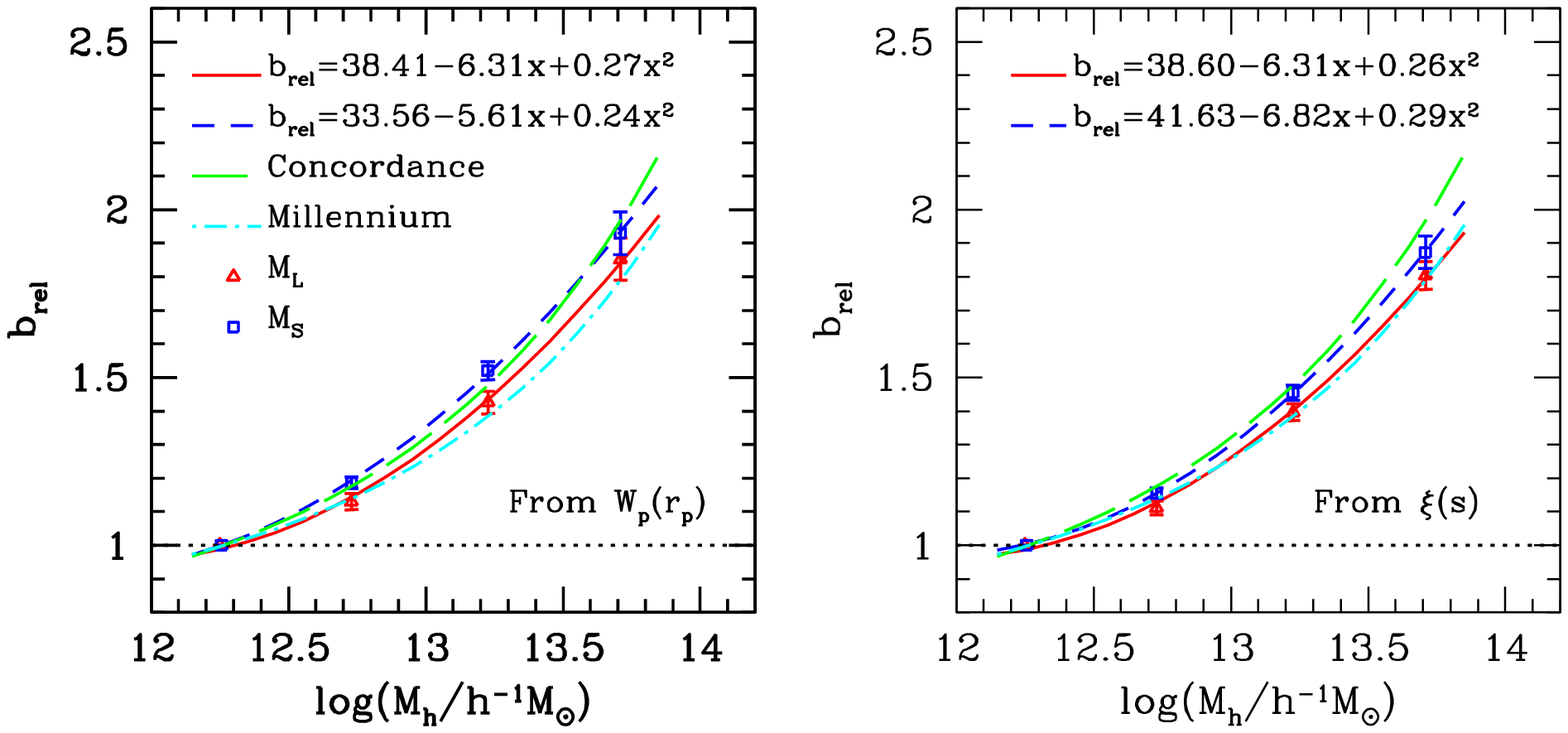} \caption{Left  panel:  The average  relative
    bias parameters,  obtained from  the projected GGCCFs  in the  range $4.19
    \mpch  \leq r_p  \leq  19.17 \mpch$  shown  in the  upper  panels of  Fig.
    \ref{fig:xi_obs}, as a function of  group mass $M_L$ (triangles) and $M_S$
    (squares).  The error bars are  measured based on the 1-$\sigma$ variances
    of 100 bootstrap re-samplings.  The relative bias parameters are normalized
    by the projected GGCCF of the  group sample G1.  The solid curve shows the
    best fitting to the relative biases of groups with mass $M_L$ (triangles).
    While the dashed curve is the best fitting relative biases for groups with
    mass $M_S$  (squares).  Right panel: The  same as the left  panel, but for
    relative bias parameters obtained using the GGCCFs in redshift space, i.e.
    the data  shown in the lower  panels of Fig \ref{fig:xi_obs}  in the range
    $4.64  \mpch \leq  s \leq  22.98 \mpch$.   The corresponding  best fitting
    results are plotted  as the solid and dashed  lines respectively.  We also
    show the relative biases obtained  from the halo model predictions (Sheth,
    Mo \& Tormen 2001) for the cosmologies we adopted for the SDSS observation
    $[\Omega_m  ,  \Omega_\Lambda   ,  \sigma_8  ]=[0.24,0.76,0.77]$  and  the
    Millennium   simulations   $[\Omega_m    ,   \Omega_\Lambda   ,   \sigma_8
    ]=[0.25,0.75,0.9]$ respectively using long-dashed and dot-dashed lines.}
    \label{fig:bias_obs_m}
\end{figure*}

In  this paper,  we use  the group-galaxy  cross  correlation function
(hereafter  GGCCF) to  study the  clustering of  galaxy groups  in the
galaxy density field.  We choose to use the  cross correlation instead
of the  auto-correlation, because in  this case much larger  number of
galaxies (compared to the number of  groups) can be used as tracers of
the underlying  density field, allowing a  more accurate determination
of the  correlation strength of the  groups. We estimate  the GGCCF in
redshift  space using  the  definition proposed  in  Davis \&  Peebles
(1983),
\begin{equation}\label{eq:xi}
\xi(s)={CG (s)~n_R \over CR(s)~n_G}-1 \,,
\end{equation} where $CG$  is the count of group-galaxy  pairs between a group
sample  and   the  corresponding   galaxy  sample;  $CR$   is  the   count  of
group-`random'  pairs between  a  group sample  and  the corresponding  random
`galaxy' samples; $s$ is the separation  of the pairs in redshift space; $n_G$
and $n_R$ are the number densities of the galaxy sample and the random sample,
respectively.  Group-galaxy and group-random  pairs are counted in logarithmic
bins  in $s$,  with bin  width $\Delta\log_{10}(s)=0.14$.
Note  that the redshift-space correlation function  is not
linearly proportional to the bias  of  the  groups. Indeed,
the  large-scale  redshift-space  mono-pole  of the auto
correlation function is
\begin{equation}
\xi(s) = \xi(r)* f(\beta) = \xi(r)*(1+2/3\beta + \beta^2/5)
\end{equation}
where $\beta=\Omega_m^{0.6}/b$. So  the amplitude  of $\xi(s)$
increases less rapidly with  halo mass than $\xi(r)$.  Assuming
$\Omega_m=0.25$, the difference in $f(\beta)$  between $b=1$ and $b=2$
is 1.33 to  1.15, respectively. Thus,  we caution  that the
relative bias measured  from  the $\xi(s)$  may  be underestimated
by a few percent ($\la 1 - \sqrt{1.15/1.33}$).

We also estimate the projected GGCCF, defined as
\begin{equation}\label{eq:Wpgg}
W(r_p)= 2\int_0^{\infty}  \xi  (r_p,r_\pi)d  r_\pi
      = 2\sum_{k}  \xi  (r_p,r_{\pi,k}) \Delta r_{\pi, k} \,.
\end{equation}
where $\xi  (r_p, r_{\pi})$ is the  GGCCF defined in a  similar way to
Eq.  (\ref{eq:xi}),   but  as  a   function  of  the   separations  of
group-galaxy pairs perpendicular ($r_p$) and parallel ($r_\pi$) to the
line-of-sight. Here,  group-galaxy and group-random  pairs are counted
in    logarithmic    bins    in    $r_p$,    with    a    bin    width
$\Delta\log_{10}(r_p)=0.17$, and in linear bins of $r_\pi$, with a bin
width   $\Delta  r_\pi=1\mpch$.    In  practice,   the   summation  in
(\ref{eq:Wpgg}) is carried over $k$  from 1 to 40, corresponding to $0
\le r_\pi \leq 40\mpch$.

Since we have two galaxy samples V1 and V2, we can make different combinations
of the group and galaxy samples so  that the depths of the samples to be cross
correlated  match  the best.   Unless  stated otherwise,  we  will  use V1  to
cross-correlate with the group samples G1 and  G2 (or S1 and S2), while we use
V2 to cross-correlate with the group samples G3 and G4 (or S3 and S4).

We  estimate the errors  of the  GGCCF using  the bootstrap  resampling method
(Barrow, Bhavsar, \& Sonoda 1984; Mo,  Jing \& B\"orner 1992).  We compute the
GGCCFs for 100  bootstrap samples of both galaxies and  groups (either real or
mock), and estimate  the errors from the scatter among  these GGCCFs. For this
purpose,  we  measure  the covariance  matrix  of  the  GGCCF, which  is  then
diagonalized.   Throughout the  paper we  use  the diagonalized  terms of  the
covariance matrix to plot the error bars.

\section{Results from the Observational Samples}
\label{sec:obs}

\subsection{The Dependence of the Group-Galaxy Cross-Correlation
on Group Mass}
\label{sec:mass_obs}

We start by  examining how the group-galaxy  cross-correlation depends
on  group  mass.   To  this   extent, we  measure  the  projected  and
redshift-space   GGCCFs     between  group  samples     G1-G4 and  the
volume-limited galaxy sample V2.   The corresponding results are shown
in Fig.  \ref{fig:xi_obs}. Upper  and  lower panels correspond to  the
projected GGCCFs and redshift-space GGCCFs, respectively.  Results are
shown for  both group mass estimates, $M_L$  and  $M_S$, separately in
the left and right-hand panels,  respectively. Clearly, the GGCCFs for
more massive groups have higher  amplitudes, although their shapes are
similar on large scales. Since the same  galaxy sample is used for all
these GGCCFs, this implies that more massive  groups are more strongly
correlated in  the galaxy density field.   Note also  that the results
for   $M_L$ and $M_S$  are very  similar, indicating  that they do not
depend strongly in which mass indicator we use.

To quantify the mass-dependence of the cross correlation amplitude, we
measure the relative  clustering bias for groups of  different masses.
We obtain the  relative bias using the average ratio  of the GGCCFs in
the range $4.19\mpch \leq r_p \leq  19.17 \mpch$ or $4.64 \mpch \leq s
\leq  22.98  \mpch$ between  group  samples  Gi  (i=1,4) and  G1.   In
Fig.~\ref{fig:bias_obs_m}, we show the relative bias measured from the
projected (left panel) and the redshift-space (right panel) GGCCFs, as
function of halo  mass.  The increase of the bias  with group mass can
be well modeled with a quadratic function, which is also shown in the
figure.   We  compare the  observed  bias  -  mass relation  with  the
prediction  of the  halo bias  model of  Sheth, Mo  \&  Tormen (2001),
adopting the same  cosmology as used in the  construction of the group
catalogue (Yang \etal  2007) (the long-dashed line), and adopting the
same cosmology as used  in the `Millennium Run' simulation (dot-dashed
line).  As  one can see, the  observed relation is  well reproduced by
the model,  but it does not  allow us to discriminate  between these two
cosmological models. These results are all in excellent agreement with
those obtained by Yang \etal (2005b) based on the groups selected from
the 2dFGRS.

\subsection{The Dependence on Group Color}
\label{sec:color_obs}

\begin{figure} \plotone{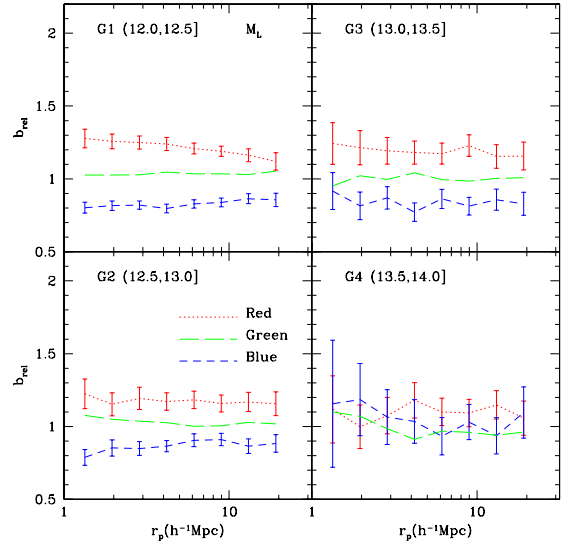} \caption{The  relative biases $b_{\rm rel}$ as
    a function  of $r_p$ obtained from  the ratios of the  projected GGCCFs of
    groups with red, green and blue  central galaxies to that of all groups in
    different group mass  bins as indicated.  In this  plot, the group masses,
    $M_L$,  are  estimated  using  the  ranking of  the  characteristic  group
    luminosity. }
  \label{fig:bias_obs_rp_L}
\end{figure}

\begin{figure}     \plotone{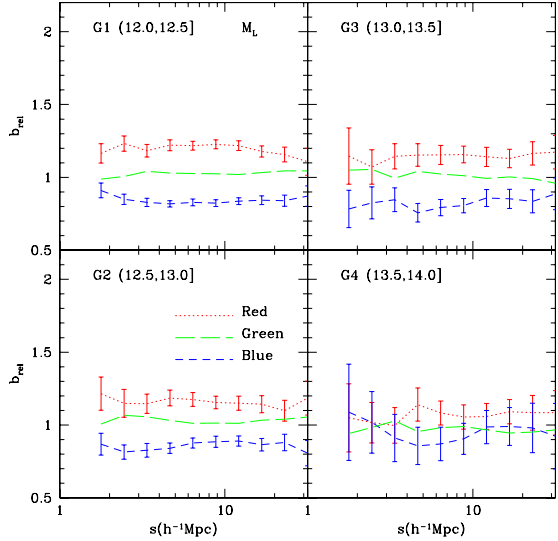}      \caption{The     same     as     Fig.
    \ref{fig:bias_obs_rp_L}, except that the relative biases $b_{\rm rel}$ are
    obtained from the redshift space GGCCFs and shown as function of $s$.}
  \label{fig:bias_obs_s_L}
\end{figure}

\begin{figure}       \plotone{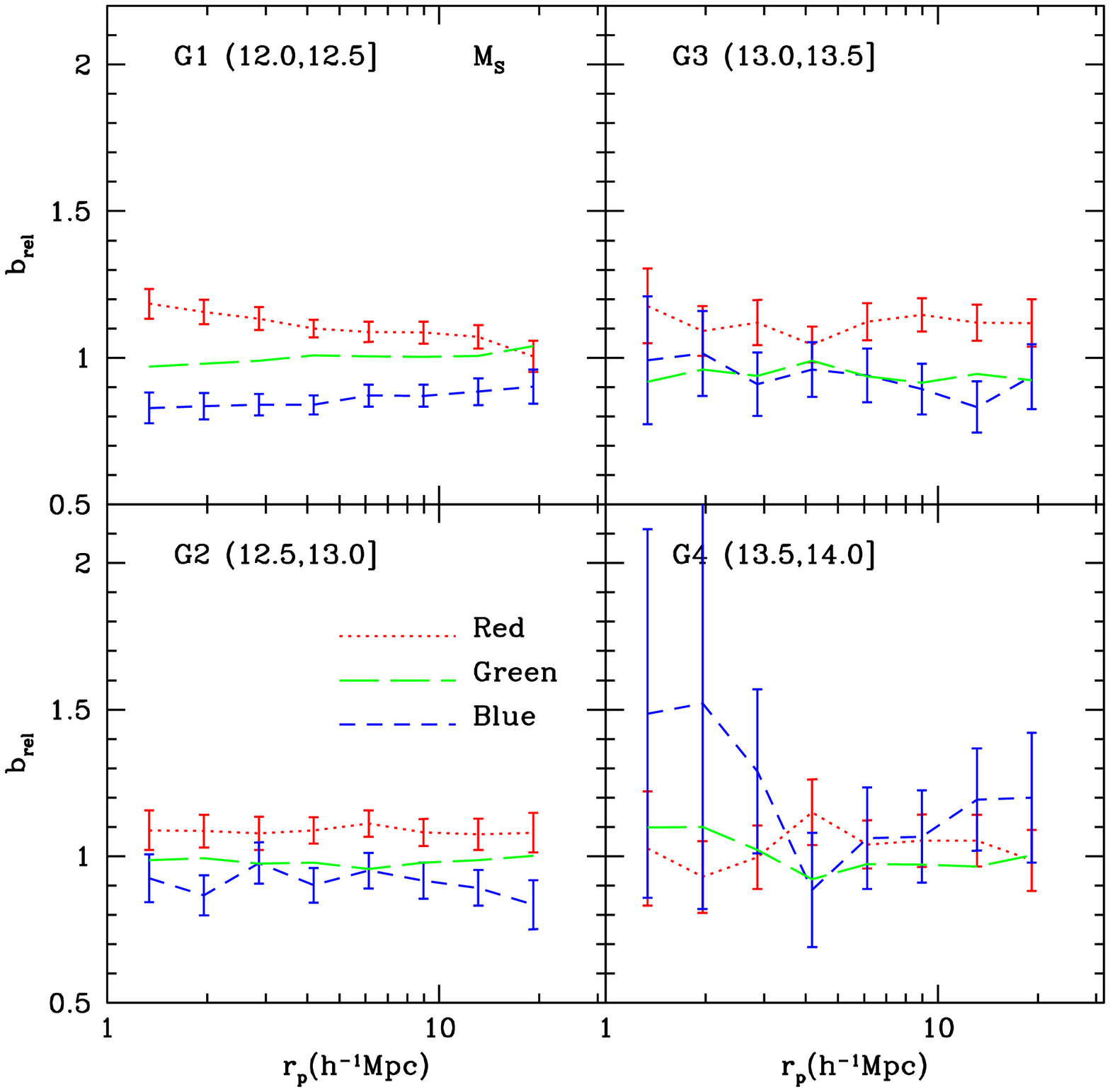}      \caption{Similar       to      Fig.
    \ref{fig:bias_obs_rp_L}, but for groups with masses, $M_S$, estimated from
    the ranking of the characteristic stellar masses of the groups. }
  \label{fig:bias_obs_rp_S}
\end{figure}

\begin{figure}       \plotone{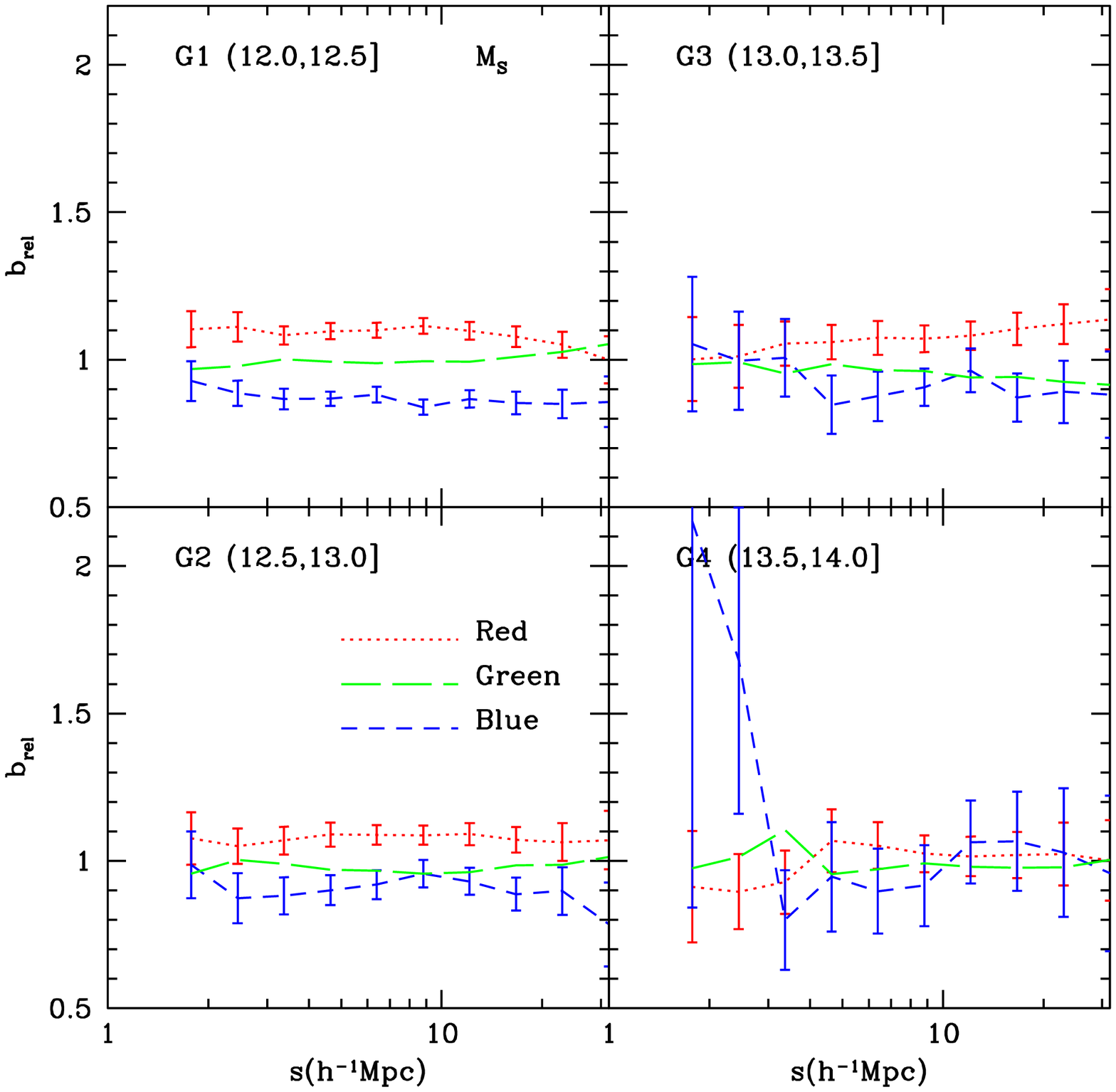}      \caption{Similar       to      Fig.
    \ref{fig:bias_obs_s_L}, but for groups  with masses, $M_S$, estimated from
    the ranking of the characteristic stellar masses of the groups.}
  \label{fig:bias_obs_s_S}
\end{figure}

\begin{figure*}  \plotone{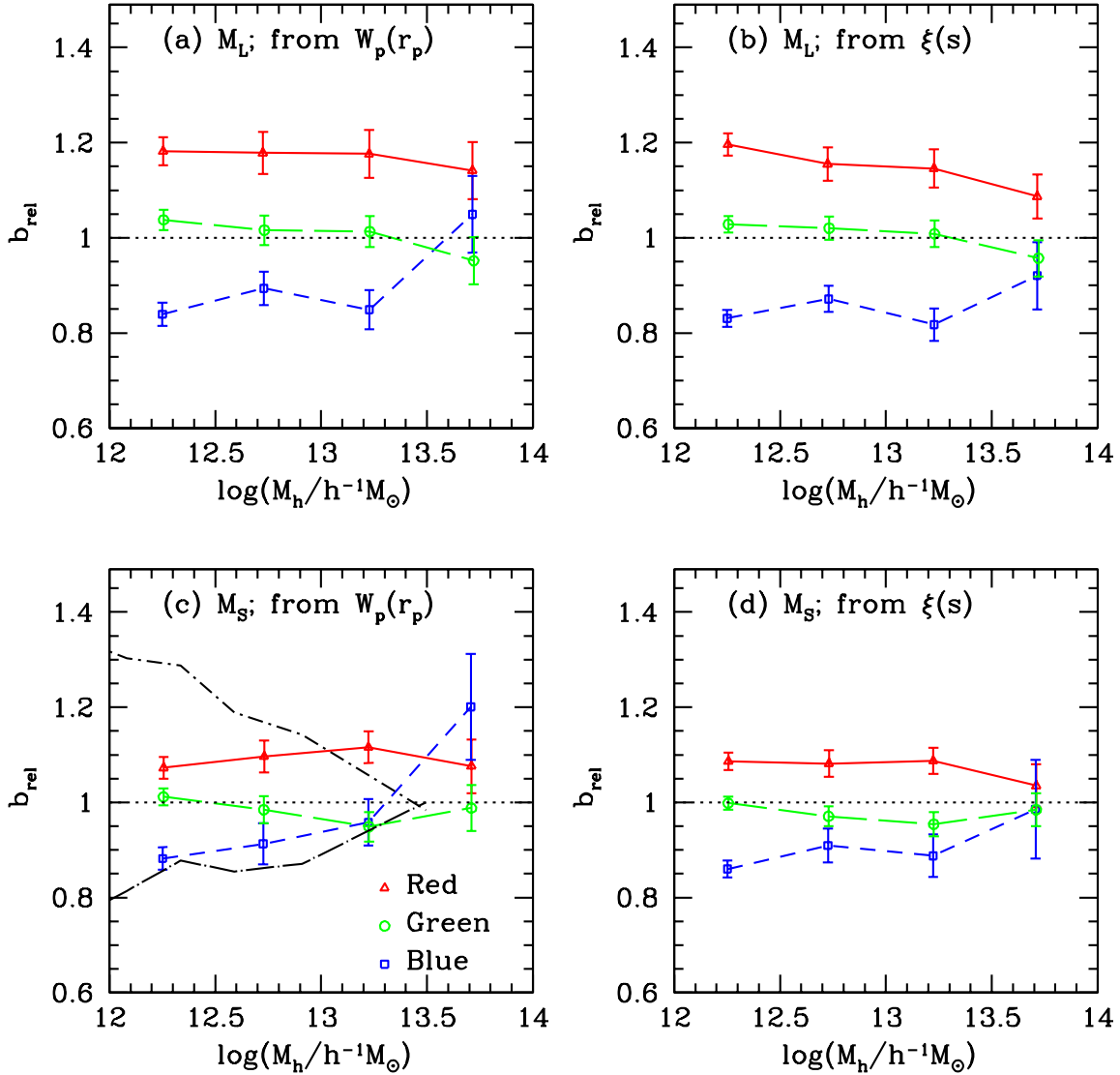}  \caption{Panel  (a): the  average  relative
    biases as a function of halo  mass (for $M_L$) obtained from the projected
    GGCCFs  shown in Fig.   \ref{fig:bias_obs_rp_L} in  the range  $4.19 \mpch
    \leq r_p \leq 19.17\mpch$. Panel (c): similar to panel (a), but for groups
    with  halo masses  $M_S$.  Panel  (b): the  average relative  biases  as a
    function of halo mass (for  $M_L$) obtained from the redshift space GGCCFs
    shown in Fig.  \ref{fig:bias_obs_s_L} in the range $4.64 \mpch \leq s \leq
    22.98 \mpch$.  Panel  (d): similar to panel (b), but  for groups with halo
    masses $M_S$.  In each panel, we  show the relative biases for groups with
    red, green  and blue  central galaxies using  open triangles,  squares and
    circles, respectively. For comparison we show also in the lower-left panel
    the age  dependence of the halo  clustering obtained by Gao  et al. (2005)
    from  N-body  simulations.    The  dot-dashed  and  dot-long-dashed  lines
    correspond  to relative  biases  for the  20\%  old and  young halos  with
    respective to all halo populations, respectively.}
\label{fig:bias_obs_colorC}
\end{figure*}

\begin{figure*}  \plotone{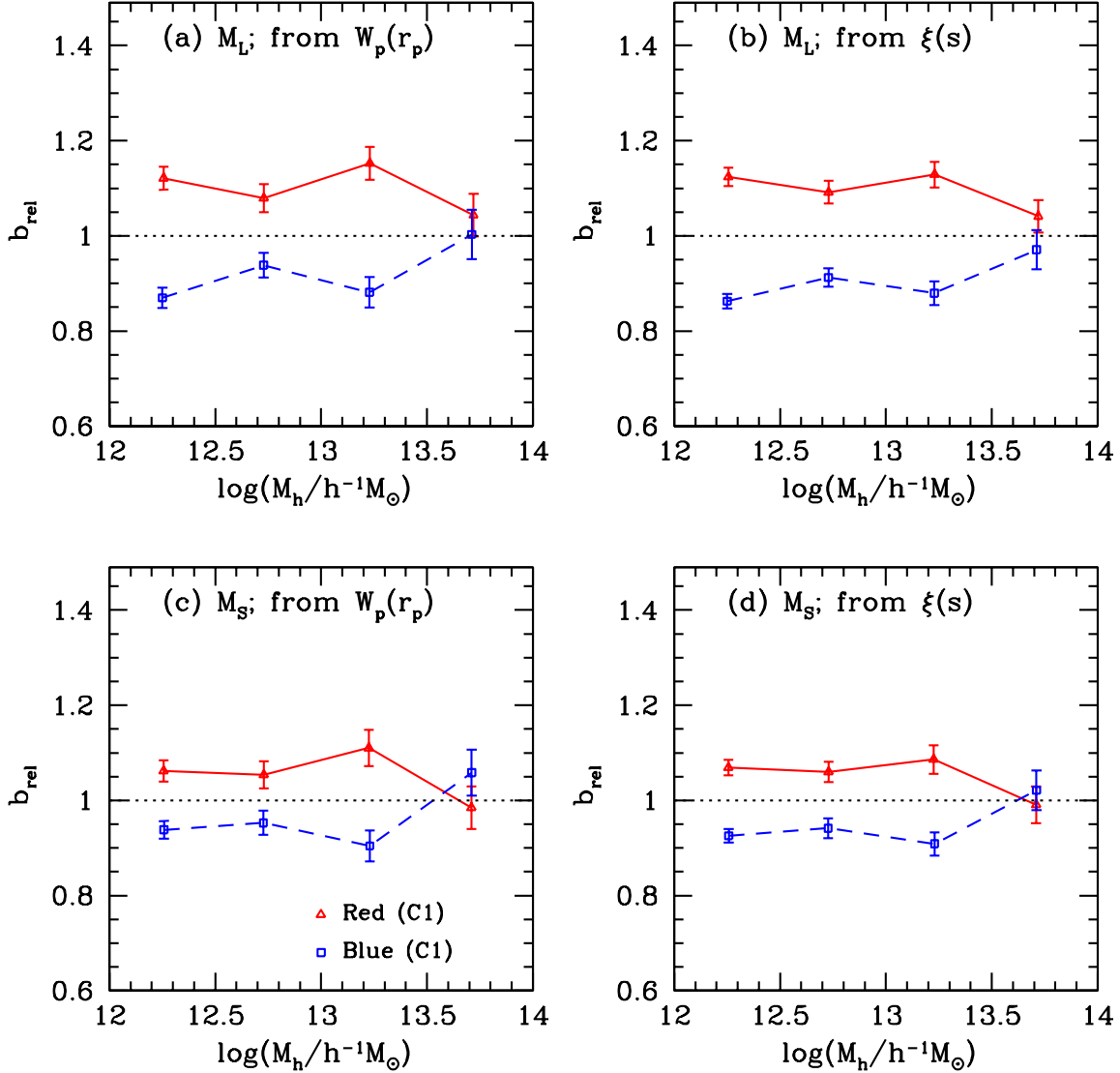}  \caption{Panel  (a): The  average  relative
    bias, obtained  using the projected GGCCFs  in the range  $4.19 \mpch \leq
    r_p \leq  19.17 \mpch$ for red  and blue groups in  terms of C1  and C2 as
    indicated.   In  this  panel  the  groups  masses,  $M_L$,  are  estimated
    according to  the ranking of  characteristic group luminosity.   The error
    bars  are  estimated  from  1-$\sigma$  variances  of  the  100  bootstrap
    re-samplings.   The relative  bias parameters  of red  and blue  groups are
    normalized by the projected GGCCFs of all groups in the corresponding mass
    bin. Panel (b): Similar to panel (a) but for relative biases measured from
    the  redshift space  GGCCFs in  the range  $4.64 \mpch  \leq s  \leq 22.98
    \mpch$.  Panels (c) and (d): Similar to panels (a) and (b), but for groups
    with masses,  $M_S$, estimated using  the ranking of  characteristic group
    stellar masses. }
  \label{fig:bias_c12}
\end{figure*}

\begin{figure*}  \plotone{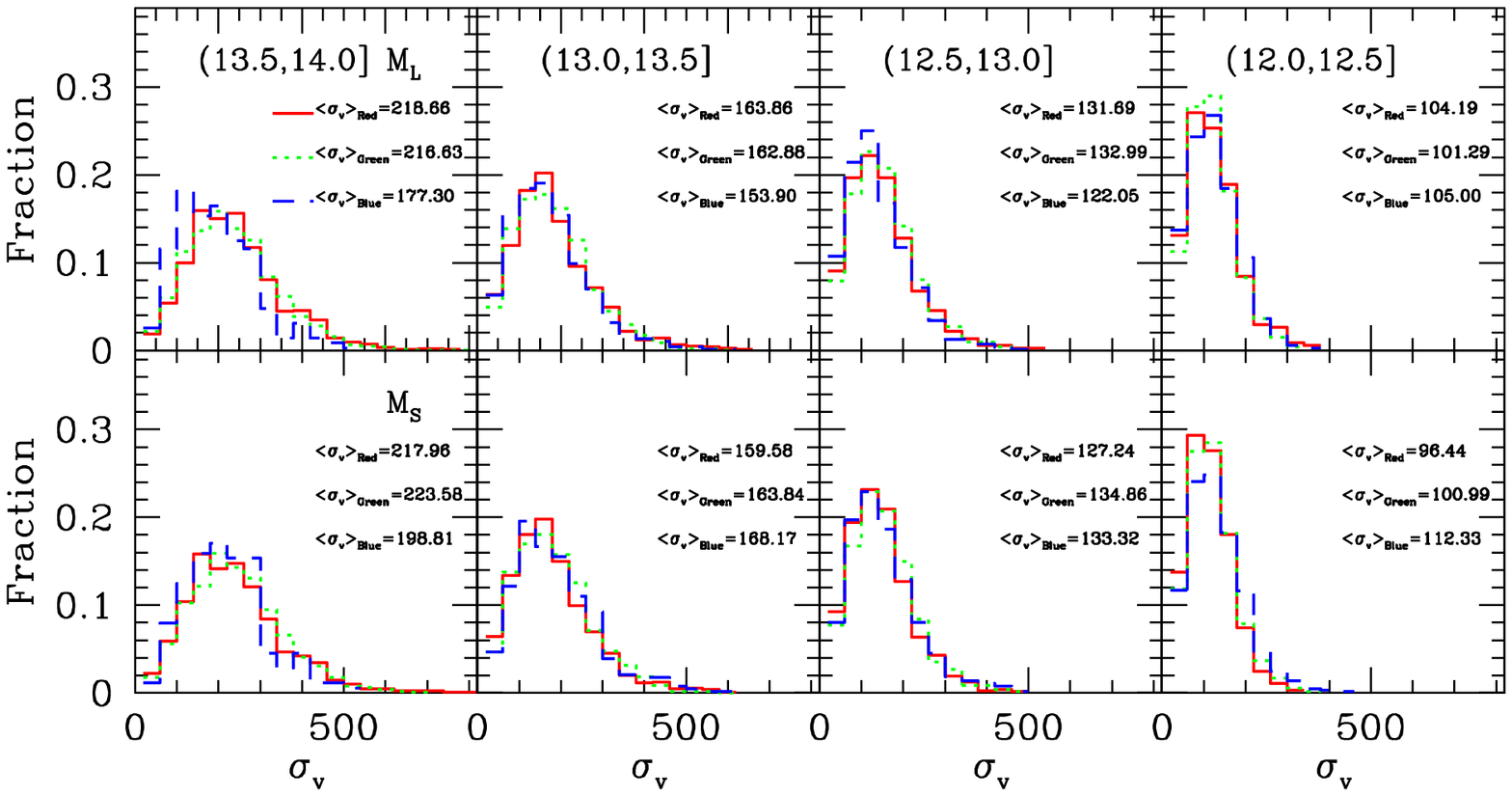}
  \caption{The distribution of the velocity dispersions for groups with red,
    blue and green centrals as indicated. The panels from left to right
    correspond to groups of different masses.  The upper and lower panels show
    the results with $M_L$ and $M_S$ as the group mass, respectively.  The
    velocity dispersions are estimated for groups with at least two members.
  }
  \label{fig:vd}
\end{figure*}

Next we investigate how the GGCCF depends on the colors of the galaxies in the
groups.  We first consider the case  C0 described above, in which the color of
a  group is  defined  by the  color of  its  central galaxy.   We measure  the
projected GGCCFs between the groups with  all, red, green or blue centrals and
the galaxy samples V1 or V2.  The relative bias for each case is then obtained
from the ratio between the projected GGCCF of the case in question and that of
the groups with all centrals.   The resulting relative bias, $b_{\rm rel}$, is
shown  as a  function of  $r_p$  in Fig.   \ref{fig:bias_obs_rp_L}.  The  four
different panels  correspond to four  different group mass  bins, G1 -  G4, as
indicated, where group masses are based on $M_L$.  The relative bias relations
for  groups  with  red,  green  and  blue  centrals  are  plotted  as  dotted,
long-dashed  and  dashed  lines,  respectively.   In order  to  obtain  better
statistics, the projected  GGCCFs are measured between the  following pairs of
group-galaxy samples:  G1-V1; G2-V1;  G3-V2 and G4-V2.   Note that  in general
groups with red  centrals are more strongly clustered than  groups of the same
mass  but with  blue centrals.   The effect  is somewhat  stronger  at smaller
separations and for  less massive groups.  Groups with  green centrals show no
significant bias relative to the total population.

We can  also measure  the relative  bias using the  GGCCFs in  redshift space.
Fig.~\ref{fig:bias_obs_s_L}  shows this  relative  bias, $b_{\rm  rel}$, as  a
function of  the redshift-space separation  $s$.  The overall behavior  of the
color dependence  of the relative bias  obtained here is very  similar to that
shown in Fig.~\ref{fig:bias_obs_rp_L}.

Note   that  the  results   presented  in   Figs.~\ref{fig:bias_obs_rp_L}  and
\ref{fig:bias_obs_s_L}  are based  on the  halo  mass estimate  $M_L$.  It  is
interesting to compare these with the results obtained using the mass estimate
$M_S$.  Since  red groups have larger  stellar masses than blue  groups of the
same luminosity, these  two estimates of halo masses, $M_L$  and $M_S$, may be
systematically different for red and blue groups of the same luminosity, which
may  affect   the  color  dependence   of  the  GGCCF  discussed   above.   In
Figs.~\ref{fig:bias_obs_rp_S} and \ref{fig:bias_obs_s_S}  we show the relative
bias of groups of different  color properties against the projected separation
$r_p$ and  redshift-space separation  $s$ for groups  with similar  $M_S$. The
color  dependence of  $b_{\rm rel}$  is  clearly significant  for groups  with
$M_S\la   10^{13.5}\msunh$,  though   it  is   weaker  than   that   shown  in
Figs.~\ref{fig:bias_obs_rp_L}   and  \ref{fig:bias_obs_s_L}.    This   can  be
explained as follows.  Because $M_L$  is smaller than $M_S$ for redder groups,
for  a given mass  bin, the  average value  of $M_S$  of the  redder subsample
defined by $M_L$ is larger than that defined by $M_S$.  Consequently, the bias
of  the  redder subsample  is  boosted  in the  $M_L$  sample  because of  the
increased halo  masses.  Although $M_S$ is  arguably a better  estimate of the
halo mass (see Yang \etal 2007), the difference in the color dependence of the
bias due  to the use  of different halo  mass estimates suggests  that further
tests are required  to assess the reliability of our  results.  We will return
to this in the next section.

To quantify the dependence of the  relative bias $b_{\rm rel}$ on the color of
the central galaxy,  we calculate the average relative  bias factor for groups
in each mass bin based on the  projected GGCCF over the range $4.19 \mpch \leq
r_p  \leq  19.17  \mpch$, and  the  corresponding  bias  factor based  on  the
redshift-space GGCCF over the range  $4.64 \mpch \leq s \leq 22.98\mpch$.  The
results are shown in Fig.\ref{fig:bias_obs_colorC}.  The upper two panels show
the relative  biases for  groups with  halo masses based  on $M_L$,  while the
lower two  panels are those for groups  with halo masses based  on $M_S$.  The
left  two panels are  results obtained  from the  projected GGCCFs,  while the
right panels are results obtained from the redshift-space GGCCFs.  The related
error bars  are estimated from the  1-$\sigma$ variances of  the 100 bootstrap
re-samplings. Note  that the  average masses of  groups are  slightly different
when the group samples G1 - G4 are divided into color subsamples. To take care
of  the contamination  due  to such  mass  difference, we  have corrected  the
relative bias of the color subsamples  relative to the whole sample, using the
mean  relation shown  in Fig.  \ref{fig:bias_obs_m} according  to  the average
$\log M_h$ of the subsample in question. Such correction is applied throughout
when we examine the color dependence of the relative bias.

As one can see from Fig.~\ref{fig:bias_obs_colorC}, at a fixed group mass, the
relative bias of  groups with red centrals is higher than  that of groups with
blue centrals.  The  color dependence is more significant  for groups of lower
masses,   and  becomes   insignificant   in  massive   groups   with  $M   \ga
10^{13.5}\msunh$.  As already  eluded to above, the results  obtained from the
projected  and  redshift-space  GGCCFs  are  very  similar,  while  the  color
dependence is weaker when using $M_S$ to estimate group masses than when using
$M_L$.  Note that  as we  will show  in  the next  section part  of the  color
dependence is  induced by systematic error,  which is stronger  for group mass
$M_L$.  To  illustrate  to  what  extent the  color-dependence  of  the  group
clustering  can be  compared with  the age-dependence  of halo  clustering, we
show, as the dot-dashed and dot-long-dashed  lines, in the lower left panel of
Fig. \ref{fig:bias_obs_colorC}  the results for  the 20\% oldest and  the 20\%
youngest halos  obtained by Gao et  al.  (2005).  The  color-dependence of the
group clustering  is qualitatively  similar to the  age-dependence of  the halo
clustering.   Quantitatively, however,  while groups  with blue  centrals have
relative bias  similar to that  of young halos,  the bias for  low-mass groups
with red  central galaxies  is significantly lower  than that of  low-mass old
halos. Thus, the  color of the central galaxy and the  assembly history of the
host halo is correlated, but the relation contains a large random component.

Following  Berlind \etal  (2006a), we  also  determine the  dependence of  the
relative bias on  the overall group color (using the  C1 subsamples defined in
section~\ref{sec_groupsamples}).  The average relative  biases of red and blue
groups thus  defined are plotted  in Fig.~\ref{fig:bias_c12} as a  function of
halo  mass\footnote{For  completeness, we  have  also  determine the  relative
biases using  the {\it average} colors  of the group members,  rather than the
total colors,  which yields results that are  virtually indistinguishable}.  A
comparison with Fig.~\ref{fig:bias_obs_colorC} shows that the color dependence
of the relative  bias based on C1 is  weaker than that based on  C0.  The main
reason  for  this  is  that  in  C0 groups  are  separated  into  three  color
subsamples, while only  two subsamples are used in  C1.  Consequently, the red
and blue subsamples  in C0 are further separated in  color-space than those in
C1 (cf. Table~\ref{tab2}).

All these results regarding the color dependence of the halo bias are in good,
qualitative agreement with those obtained by Yang \etal (2006) from the
2dFGRS, as long as red galaxies are mainly the galaxies with passive star
formation.  However, they are completely opposite to the results obtained by
Berlind \etal (2006a) who, using the SDSS group catalogue of Berlind \etal
(2006b), found that (i) blue groups are more strongly clustered than red
groups of the same mass, and (ii) the color dependence is stronger for more
massive groups.  Since the galaxy samples and statistical methods used are
very similar, these differences more likely result from differences in the
group-finding algorithms used or the groups selected.  The general properties
of the groups selected by Berlind et al.  are similar to the groups used
here. For instance, relationship between group richness and halo mass is about
the same for both group catalogs. However, we do find that the central galaxy
luminosity - halo mass relation in their group samples is quite different from
that in ours, particularly for groups with masses $\la 10^{13.5}\msunh$.  In
their Fig.\,3(d), Berlind et al. (2006a) showed that the central galaxies in
low-mass halos have three or four distinctive populations (with unfilled
gaps), which are not present in our samples. Thus the groups, especially the
low-mass ones, are somewhat different in the two catalogs.  On the other hand,
as we will show in the next section using mock samples, the use of $M_L$ as
group mass induces systematic errors that can produce a significant part of
the observed color dependence of the clustering bias factor.  Since Berlind et
al. have also used the group masses estimated using the total group
luminosities, their results might also be affected by such systematic
effect. Unfortunately, the bias due to the systematic effect is expected to be
in the opposite direction as color-dependence found by Berlind et al.
(2006a), namely, it should enhance the clustering amplitude of red
groups. While it is beyond the scope of our paper to identify in detail the
origin of the discrepancy, we use mock samples in the next section to test the
reliability of our results.

Before proceeding to the next section, we follow Berlind \etal (2006a) to
check whether the velocity dispersions of the groups in halos of the same
assigned mass but with centrals of different colors are the same. This
provides yet another test about the reliability of our mass assignment. Here
we measure the velocity dispersions for groups with at least two members using
the gapper estimator described by Beers, Flynn \& Gebhardt (1990; see also
Yang et al. 2005a).  Although the estimate of the velocity dispersion is
unreliable for estimating individual halo mass, the stacking result may be a
reliable indicator of the mean mass of a set of groups. To show that the
color-dependent bias is not a result of systematic mass difference between the
color subsamples, we estimate the mean velocity dispersion of each color
subsample. The results are shown in Fig.\,\ref{fig:vd} for groups of different
assigned masses, with upper and lower panels corresponding to masses based on
$M_L$ and $M_S$, respectively. The results show clearly that there is no
systematic difference between the mean velocity dispersions of groups with
red, blue and green centrals.  Thus, the color dependence we found in the SDSS
observation is unlikely due to systematics in the mass assignments.


\begin{figure*}
  \plotone{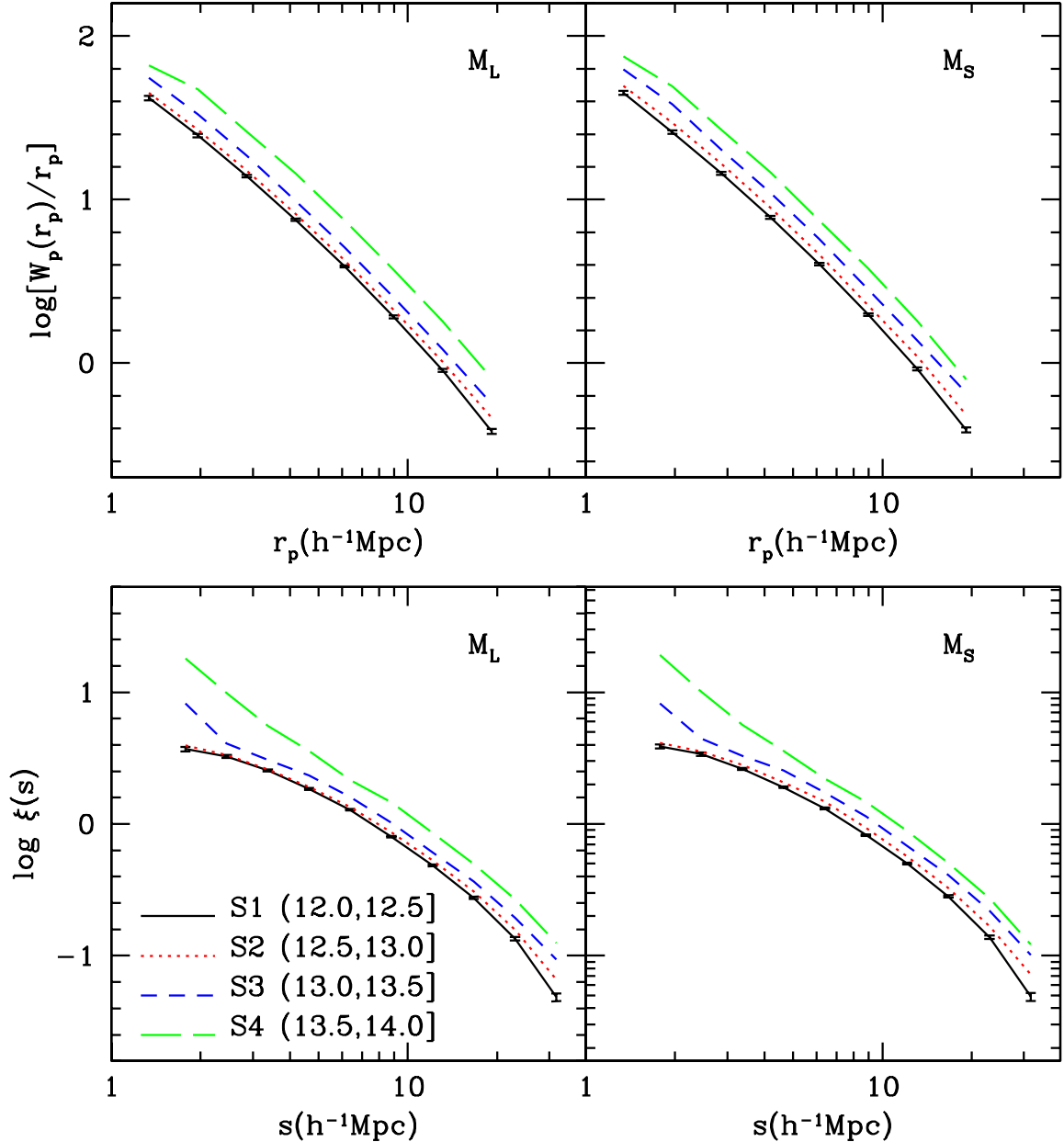} \caption{Similar  to Fig. \ref{fig:xi_obs}, except
    that the results are for  the MGRS constructed from the Millennium
    galaxy catalogue. }
  \label{fig:xi_mgrs}
\end{figure*}

\begin{figure*}
  \plotone{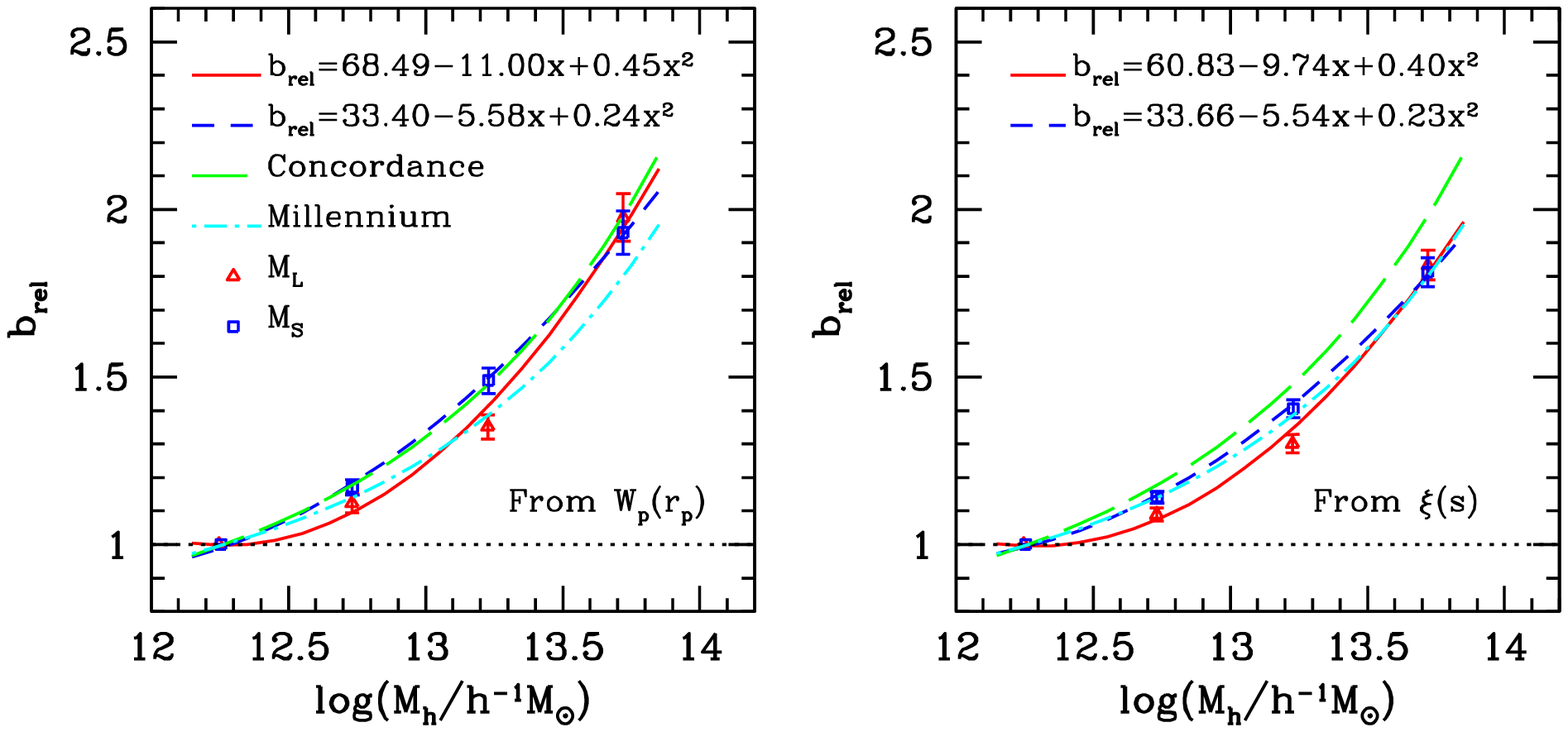}  \caption{Similar   to  Fig. \ref{fig:bias_obs_m},
    except that the results are   for  the MGRS constructed from   the
    Millennium galaxy catalogue.}
  \label{fig:bias_mgrs_m}
\end{figure*}

\begin{figure*}      \plotone{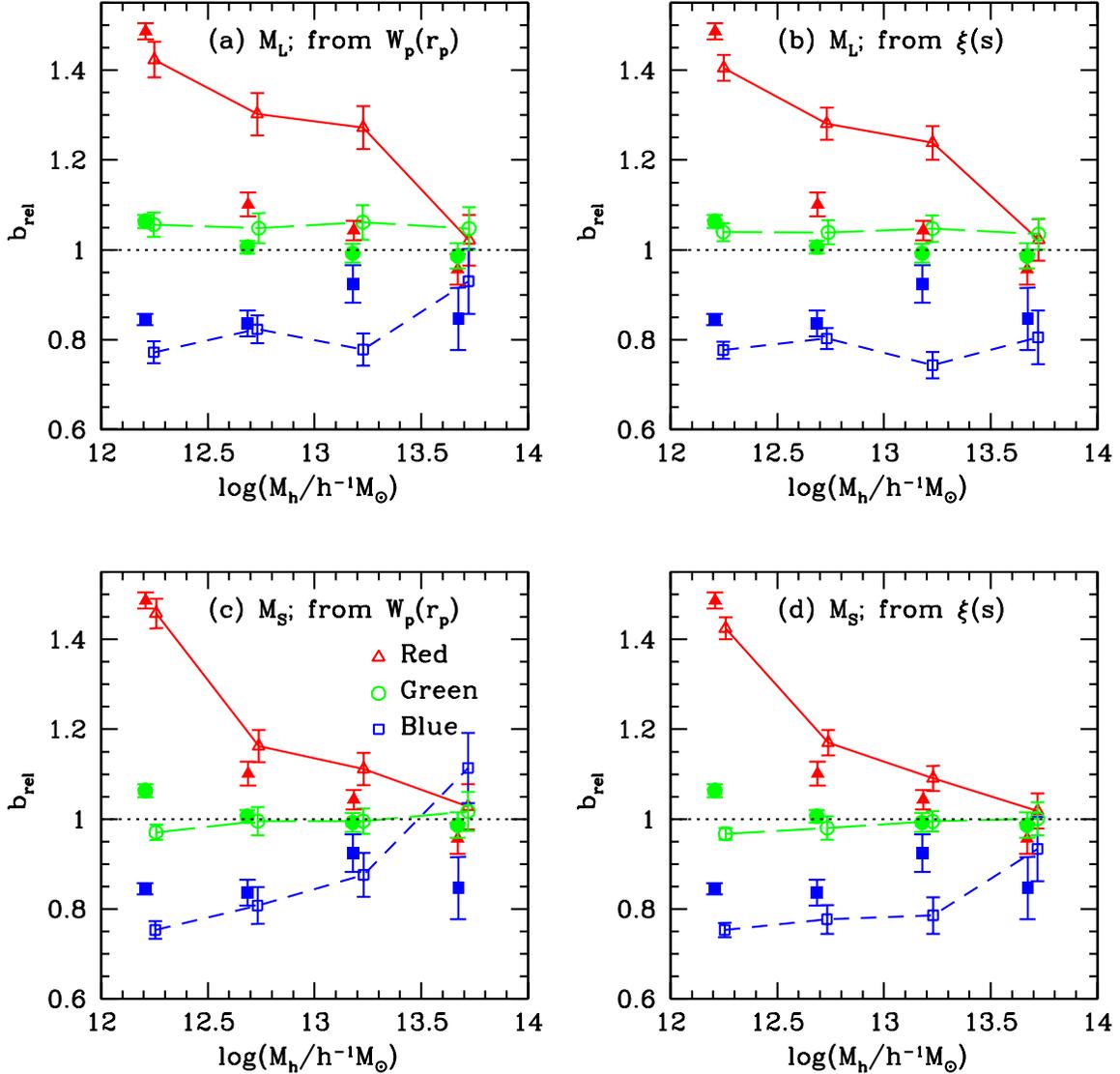}      \caption{Similar      to      Fig.
    \ref{fig:bias_obs_colorC}, but  for results  obtained from the  MGRS (open
    symbols  with lines  connected).   Note that,  for  comparison, the  solid
    symbols  in  this  plot  are  results obtained  from  the  {\it  original}
    Millennium simulation galaxy and halo  catalogues, and the real space auto
    correlation  functions   are  used  in  measuring   the  average  relative
    biases. For clarity,  the latter set of data are  slightly shifted to left
    hand side. }
    \label{fig:sam_bias_c0}
\end{figure*}

\begin{figure*}      \plotone{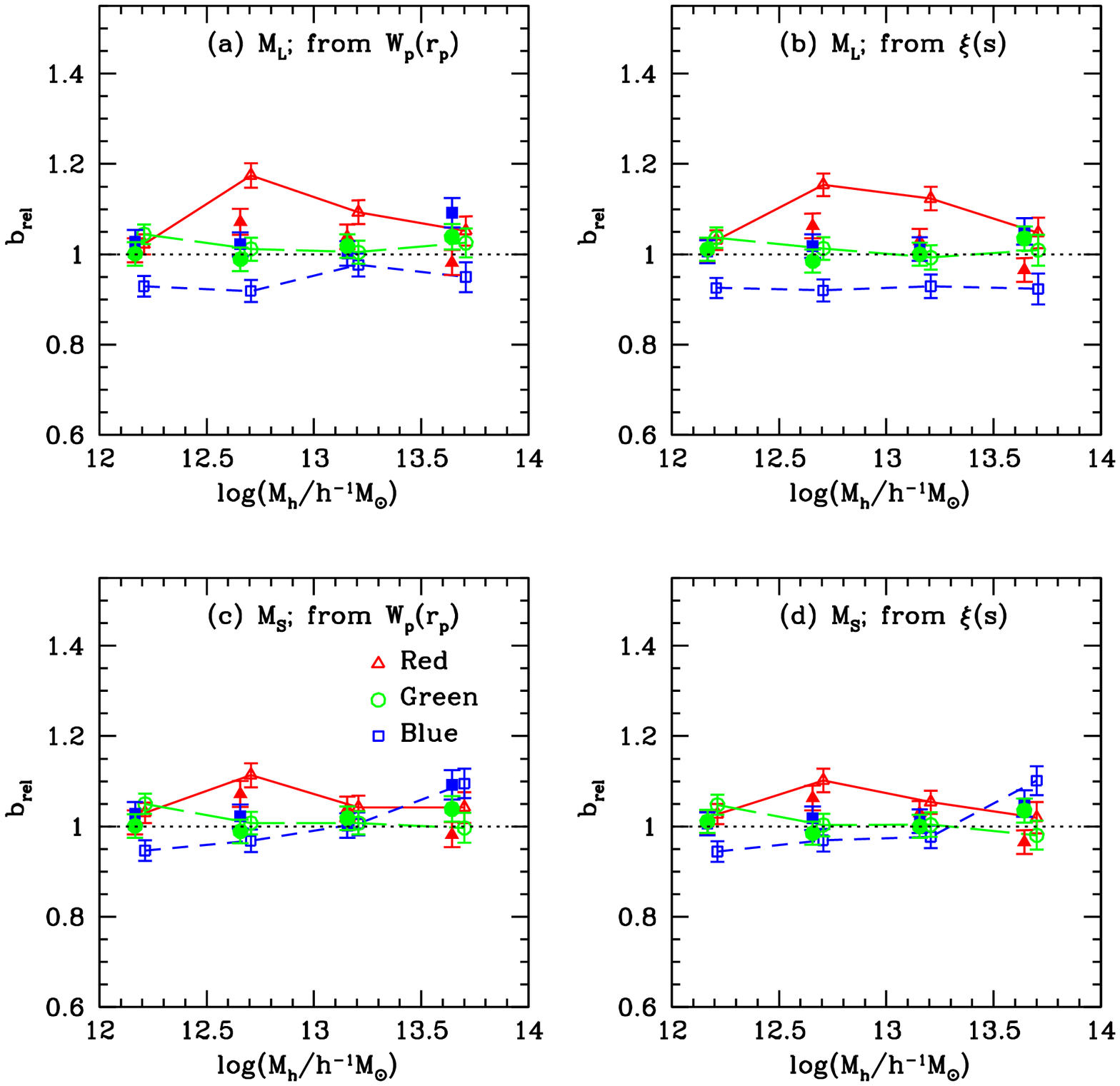}      \caption{Similar      to      Fig.
    \ref{fig:sam_bias_c0},  but  for  results  obtained  from  the  MGRS  with
    randomly  re-assigned galaxy  colors according  to the  color-stellar mass
    distributions of the central and satellite galaxies separately in the SDSS
    observations.  See text for details. }
    \label{fig:random}
\end{figure*}

\section{Results from the Mock Samples}
\label{sec:MGRS}

In  this section,  we use  the  MGRS constructed  from the  Millennium
semi-analytical  galaxy catalogue (Croton  \etal 2006)  to investigate
the  mass and  color dependence  of the  GGCCF.  The  purpose  here is
twofold.   First of  all,  as mentioned  above,  we want  to test  the
reliability of our results against uncertainties in the identification
of galaxy groups, and in the mass assignments.  Secondly, we also want
to  investigate whether  the  semi-analytical model  of Croton  \etal
which is  fairly successful  in matching observational  data, actually
predicts a  color dependence of the  halo bias and how  it compares to
the data presented here.

\subsection{Dependence on Group Mass}

The  GGCCFs for the  MGRS have  been measured  using exactly  the same
method  used in Section~\ref{sec:mass_obs}  for the  real data.   Fig.
\ref{fig:xi_mgrs} shows  the GGCCFs for mock groups  in different mass
bins.  Comparing these results with the observational results shown in
Fig.\ref{fig:xi_obs}), one sees that  the mass dependence predicted by
the  model  is   very  similar  to  that  obtained   from  the  SDSS.
Fig.~\ref{fig:bias_mgrs_m} shows the  corresponding relative bias as a
function of  group mass.  As  in the observational data,  the relative
biases are  defined with  respect to groups  with masses in  the range
$10^{12.0}-10^{12.5}\msunh$.  The mass dependence of the relative bias
can  again be  modeled  with  quadratic forms,  as  indicated in  the
panels.  As  with the SDSS data,  the results are well  matched by the
predictions  based  on  the  clustering  of dark  matter  halos  in  a
$\Lambda$CDM  cosmology, suggesting  that  our group  finder is  quite
reliable in  grouping galaxies according  to their common  dark matter
halos. In addition,  the halo masses assigned to  the groups also have
to  be fairly  reliable,  otherwise we  would  not have  been able  to
recover the strong mass dependence shown. This is also consistent with
the tests described  in Yang \etal (2007), where it  is shown that the
masses assigned to  the groups agree with the true  halo masses with a
scatter $\lta 0.3$ dex.

\subsection{The Dependence on Group Color}

Fig.\ref{fig:sam_bias_c0} shows the relative biases  of red, green and
blue groups in  the MGRS as function  of the assigned group mass. Here
the colors refer to those of the central galaxies (i.e., the brightest
group  members), which have been  determined using the color-magnitude
relation shown in the right-hand  panel of Fig.~\ref{col:gau}.  As for
the real data, the  relative bias is defined relative  to the GGCCF of
all groups   in the mass bin.   In  the left-hand panels  the relative
biases have  been estimated from the projected  GGCCFs  using the data
over the range $4.19 \mpch \leq r_p \leq 19.17 \mpch$.  The right-hand
panels     shows  the  relative    bias  factors    obtained  from the
redshift-space GGCCFs using the data  in the range  $4.64 \mpch \leq s
\leq 22.98\mpch$.  The results shown in the upper two panels use group
masses  based on $M_L$,  while the lower   two panels use group masses
based  on  $M_S$.  Compared with   the  observational results shown in
Fig.~\ref{fig:bias_obs_colorC}, the     MGRS  reveals a  significantly
stronger color dependence, especially for   groups in the lowest  mass
bin, $10^{12.0}-10^{12.5}\msunh$.

In order to investigate to what extent this color dependence of the GGCCF is
affected by selection effects, we compare the results obtained from the MGRS
with those obtained directly from the Millennium galaxy catalogue.  For this
purpose, we measure the relative biases directly from the Millennium
simulation and use the color of the central galaxy to separate halos of a
given mass into red, green and blue subsamples.  Here we use the
auto-correlation function of the dark matter halos (instead of the GGCCF) {\it
  in real space} to estimate the relative bias of halos in different
subsamples.  The relative biases are defined relative to the correlation
function of all groups in the mass bin in question, and are estimated using
data in the separation range $8.98 \mpch \leq r \leq 44.6 \mpch$.  The solid
symbols in Fig.~\ref{fig:sam_bias_c0} show the relative bias factors of red,
green and blue halos thus obtained as function of the true halo mass.
Clearly, the SAM predicts a color dependence.  Similar to our results, red
halos are more strongly clustered than blue halos of the same mass, and the
color dependence is stronger for less massive halos.  A comparison with the
corresponding open symbols, which show the results obtained from the mock
group catalogue constructed from the MGRS, indicates that the color dependence
in the original data is fairly well recovered by our analysis of the GGCCF.
In particular, when using $M_S$ as halo mass estimator, the true trends can be
recovered fairly accurately.  Quantitatively however, there is discrepancy
between the original signal in the Millennium Simulation and that recovered
from the mock groups, especially for blue groups when GGCCFs in redshift space
are used, in the sense that the recovered color dependence is stronger than
the dependence in the original data.  The discrepancy is larger when $M_L$ is
used.  The amount of systematics is comparable to the difference between red
and blue groups in the observational sample [see panels (c) and (d) of
Fig.\ref{fig:bias_obs_colorC}].  In order to check further how the color
dependence of the group clustering in the SDSS data can be affected by the
group finder used and the model of halo-mass assignments, we carry out an
additional test. Here we first randomly re-assign to each central or satellite
galaxy of a given stellar mass in the Millennium Simulation a
$^{0.1}(g-r)$-color according to the color-stellar mass relation of the
central or satellite galaxies in the SDSS observation. The absolute magnitude
of each mock galaxy is then re-calculated according to the new assigned
color. Note that such a color assignment according to the SDSS observation
makes sure that the color distribution of mock galaxies is the same as that in
the observation, thereby reducing any effects that may be produced by the
difference in the color distribution.  Thus, the re-constructed new mock
galaxy catalogue has a similar color-stellar mass relation as the SDSS
observation.  More importantly, there should not be any color-dependence of
group clustering in this new mock catalog.  Using this catalogue, we generate
the MGRS again and perform all the same analyses (i.e. finding groups,
assigning halo masses according to luminosity or stellar mass, and calculating
the group correlation function) to obtain the relative bias of corresponding
galaxy groups. The results are shown in Fig. \ref{fig:random} as the open
symbols for groups with red (triangles), green (circles) and blue (squares)
central galaxies, defined with the color criteria that is exactly the same as
in the observation. For comparison, we also show, as the solid symbols, the
true relative bias based on the original {\it halos} in the MGRS. The use of
halos in the {\it MGRS} ensures that we are comparing the {\it same} set of
halos (groups) in the same large-scale structures and the differences between
the two are simply due to the systematic errors. Ideally, the relative biases
for the halos in the MGRS should contain no color dependence, and the
deviation of the solid symbols from the null hypothesis, $b_{rel}=1$, which is
at roughly the 1-$\sigma$ level, reflects the `cosmic variance' in the color
assignment.  As can be seen, there is some difference between the open and
solid symbols, which indicates the systematic errors introduced by the
group-finding procedure and the halo-mass assignment. The systematic errors
are $\la 0.1$ for the cases using $M_L$ (see the upper two panels of
Fig. \ref{fig:random}), and are $\la 0.05$ for the cases using $M_S$ (see the
upper two panels of Fig. \ref{fig:random}).  These are significantly smaller
than the difference shown in Fig.  \ref{fig:bias_obs_colorC}. Finally we have
also made a test in which each central or satellite galaxy in the Millennium
Simulation is randomly re-assigned a $^{0.1}(g-r)$ color according to the
color-{\it magnitude} relation of the central or satellite galaxies in the
SDSS.  The overall systematics here is slightly smaller.  Based on all these
tests with the mock catalog, we conclude that our results are robust against
the systematics produced by the group-finding procedure and the halo-mass
assignment.

\section{Summary and Discussion}
\label{sec:summary}

In this paper we have used a large galaxy group catalogue constructed from the
SDSS Data Release 4 (DR4) to study how the clustering of galaxy groups depends
on their masses and, at fixed mass, on the color of their member galaxies.  We
use  the projected  and redshift-space  two-point cross  correlation functions
between  groups and  galaxies to  measure the  relative biases  for  groups of
different masses and colors.  Our main results can be summarized as follows:
\begin{enumerate}
\item The  correlation amplitude  of galaxy groups  depends strongly  on their
  masses, and  is in good  agreement with the  mass dependence of the  bias of
  dark matter halos in the $\Lambda$CDM concordance cosmology.
\item For  a given mass, the  correlation amplitude of groups  also depends on
  the  characteristic color  of  its  member galaxies.   Red  groups are  more
  strongly  clustered  than  blue  groups.   This  color  dependence  is  more
  prominent  in low-mass  groups,  and becomes  insignificant  in groups  with
  masses  $\gta  10^{14}\msunh$.   These  results  are  in  good,  qualitative
  agreement  with those obtained  by Yang  \etal (2005b)  from an  analysis of
  galaxy  groups in  the 2dFGRS,  but disagree  with the  results  obtained by
  Berlind \etal (2006a).
\item The observed color dependence in the data is qualitatively reproduced by
  the semi-analytical  model of Croton  \etal (2006), but the  predicted color
  dependence  is much  too  strong for  low-mass  halos with  $M \sim  10^{12}
  h^{-1}\Msun$.
\item The systematic errors that can be induced by the group finder or by the
  method of assigning masses to the groups are discussed. And we conclude that
  our finding of the color dependence of the relative bias in the SDSS
  observation is robust.
\end{enumerate}
Unlike   the  mass   dependence  of   the  group   clustering  which   can  be
straightforwardly  linked  to  the  mass  dependence  of  the  halo  bias,  an
interpretation of the color dependence  at fixed mass is less straightforward.
It  is tempting  to  link  it to  the  assembly bias  that  has recently  been
discovered in  numerical simulations, and which  shows that haloes  of a given
mass that assemble earlier are more strongly clustered. The assembly time of a
dark matter halo, $t_{\rm main}$, is typically defined as the lookback time at
which its  main progenitor reaches a mass  that is half of  the halo's present
day mass.

A straightforward, but naive, interpretation  of our results therefore is that
halos  that assemble  earlier contain  redder  galaxies. This  would not  only
explain the sign of the color-dependence at fixed halo mass detected here, but
since the assembly bias becomes less significant for more massive halos (e.g.,
Gao  \etal  2005;  Jing  \etal  2007),  it also  naturally  explains  why  the
color-dependence  is weaker  for more  massive groups.   One way  to  invoke a
positive,  causal correlation between  the assembly  histories of  dark matter
haloes  and the star  formation histories  of the  associated galaxies,  is to
assume that the  epoch of the last major merger,  which is strongly correlated
with the assembly  time (Li \etal 2007b), also signals the  time at which star
formation  is terminated.  This  may come  about if  the feedback  from merger
induced  AGN activity expels  (or heats)  the gas  associated with  the merger
remnant,  thus shutting  off the  star  formation activity  (e.g., Di  Matteo,
Springel \& Hernquist 2005; Hopkins  \etal 2006, 2007).  Such a merger induced
truncation  mechanism would ensure  that halos  that assemble  earlier contain
redder galaxies.

Alternatively,  the   observed  color  dependence   may  be  related   to  the
`archeological' downsizing.  It is well known that,  in hierarchical structure
formation, more  massive halos assemble later  (e.g., Lacey \&  Cole 1993; van
den  Bosch 2002;  Wechsler  \etal  2002).  Therefore,  if  indeed haloes  that
assemble earlier  contain redder  galaxies, it would  imply that  more massive
haloes (i.e., clusters) contain bluer  galaxies than low mass haloes.  This is
in clear  conflict with observations,  which show that more  massive galaxies,
which typically  reside in  more massive halos,  on average have  formed their
stars earlier.   An attractive solution to  this problem has  been proposed by
Neistein \etal  (2006), who introduced  an alternative `halo  formation time',
the  time at  which  the  sum over  the  masses of  {\it  all} the  virialized
progenitors with masses  above a given minimum mass,  reaches half the present
day halo  mass.  As shown by  Neistein \etal (2006), contrary  to the assembly
time,  $t_{\rm  main}$,  defined  as  the  accretion time  of  the  last  main
progenitor, $t_{\rm all}$  {\it increases} with increasing halo  mass, so that
it provides  a natural explanation  for the observed  archeological downsizing
that galaxies  in more massive haloes  have older stellar  populations. If the
star formation history of a galaxy  is correlated with this formation time, it
may also  explain the color-bias  at fixed halo  mass. The reason is  that, as
demonstrated in  Neistein \etal (2006),  although the formation  time, $t_{\rm
all}$, and  the assembly  time, $t_{\rm main}$,  of halos  are anti-correlated
when considering the entire halo  population, at fixed halo mass $t_{\rm all}$
and $t_{\rm main}$ are positively correlated.  Thus, at fixed halo mass, a halo
that  forms  earlier will  contain  a redder  galaxy  and  will also  assemble
earlier.


\section*{Acknowledgments}

YW and XY acknowledge the Max-Plank Institute for Astrophysics for hospitality
and financial  support where this work  was finalized.  We thank  Cheng Li for
useful discussions, Liang Gao for providing us the data in electronic form and
the  anonymous  referee  for   helpful  comments  that  greatly  improved  the
presentation of this  paper. XY is supported by the  {\it One Hundred Talents}
project,  Shanghai   Pujiang  Program   (No.  07pj14102),  973   Program  (No.
2007CB815402), the  CAS Knowledge Innovation Program  (No.  KJCX2-YW-T05), and
grants from NSFC (Nos. 10533030, 10673023).  HJM would like to acknowledge the
support  of  NSF  AST-0607535,  NASA  AISR-126270 and  NSF  IIS-0611948.   The
Millennium Run  simulation used  in this  paper was carried  out by  the Virgo
Supercomputing Consortium at the Computing Centre of the Max-Planck Society in
Garching.   The  semi-analytic  galaxy  catalogue  is  publicly  available  at
http://www.mpa-garching.mpg.de/galform/agnpaper.



\begin{thebibliography}{}

\bibitem[] {Adelman-McCarthy-06}
{Adelman-McCarthy} J.~K., {Ag{\"u}eros} M.~A., {Allam} S.~S.,
{Anderson} K.~S.~J., {Anderson} S.~F., {Annis} J.,
{Bahcall} N.~A., {Baldry} I.~K., {et al.,} 2006, \apjs, 162, 38

\bibitem[]{Bah92}
Bahcall, N. A., West, M. J., 1992, \apj , 392, 419

\bibitem[]{Bal04}
Baldry, I. K., Glazebrook, K., Brinkmann, J.,
  Ivezi$\acute{c}$, $\breve{Z}$eljko; Lupton, R. H., Nichol, R. C., Szalay, A.
  S. ,2004 , \apj , 600 ,681

\bibitem[]{Bal06}
Baldry, I. K., Balogh, M. L., Bower, R. G., Glazebrook, K., Nichol,
R. C., Bamford, S. P., Budavari, T., 2006, \mnras, 373, 469

\bibitem[]{Barr84}
Barrow, J. D., Bhavsar, S. P., Sonoda, D. H., 1984, \mnras, 210, 19


\bibitem[]{Beers90}
Beers T.C., Flynn K., Gebhardt K., 1990, \aj , 100, 32


\bibitem[]{Ber02}
Berlind, A. A., \& Weinberg, D. H. 2002, \apj , 575, 587

\bibitem[]{Bel03}
Berlind A.A., Weinberg D.H., Benson A.J., Baugh C.M., Cole
  S., Dave R., Frenk C.S., Jenkins A., Katz N., Lacey C.G., 2003, ApJ, 593, 1

\bibitem[]{Ber06a}
Berlind, A. A., Kazin, E., Blanton, M. R., Pueblas, S.,
  Scoccimarro, R., Hogg, D. W., 2006a, astro-ph/0610524

\bibitem[]{Ber06b}
Berlind, A. A. et al. 2006b, ApJS, 167, 1

\bibitem[]{Blanton-03-Kcorrection}
{Blanton} M.~R., {Brinkmann} J., {Csabai}
  I., {Doi} M., {Eisenstein} D., {Fukugita} M., {Gunn} J.~E., {Hogg} D.~W.,
  {et al.,} 2003a, \aj, 125, 2348

\bibitem[]{Blanton-03-LF}
{Blanton} M.~R., {Hogg} D.~W., {Bahcall} N.~A.,
  {Brinkmann} J., {Britton} M., {Connolly} A.~J., {Csabai} I., {Fukugita} M.,
  {et al.,} 2003b, \apj, 592, 819

\bibitem[]{}
Blanton M.R., Eisenstein D.J., Hogg D.W., Schlegel D.J.,
  Brinkmann J., 2005a, \apj, 629, 143

\bibitem[]{NYUVAGC}
Blanton M.R. \etal, 2005b, \aj, 129, 2562

\bibitem[]{Blan07}
Blanton, M. R., Berlind, A. A., 2007, \apj , 664, 791

\bibitem[]{Col05}
Collister, A. A., \& Lahav, O. 2005, \mnras , 361, 415

\bibitem[]{Coo02}
Cooray, A., \& Sheth, R. 2002, Phys. Rep. 372, 1

\bibitem[]{Coor06}
Cooray A., 2006, \mnras , 365, 842

\bibitem[]{Coll01}
Colless, M., \& The 2dFGRS team 2001, \mnras , 328, 1039

\bibitem[]{Cro06}
Croton, D. J., S., Volker, White, Simon D. M. , et al.,
  2006, \mnras ,365 ,11

\bibitem[]{Cro07}
Croton, D. J., Gao, L., White, S. D. M., 2007, \mnras , 374, 1303

\bibitem[]{}
Davis, M., Peebles, P.J.E.\ 1983, ApJ 267,465

\bibitem[]{}
{de Vaucouleurs}, G., {de Vaucouleurs}, A.,
  {Corwin}, H.~G., {Buta}, R.~J., {Paturel}, G., \& {Fouque}, P. 1991, {Third
    Reference Catalogue of Bright Galaxies} (Volume 1-3, XII, 2069 pp.~7
  figs..~ Springer-Verlag Berlin Heidelberg New York)

\bibitem[]{DiM05}
Di Matteo T., Springel V., Hernquist L., 2005, Nature, 433, 604

\bibitem[]{Gao05}
Gao, L., Springel, V., \& White, S. D. M. 2005, \mnras ,
  363, L66

\bibitem[]{}
Gao L., White S. D. M., 2007, \mnras, 377, L5


\bibitem[]{}
Hahn O., Porciani C., Carollo C.M., Dekel A., 2007, \mnras, 375, 489

\bibitem[]{Harker06}
Harker, G.; Cole, S.; Helly, J.; Frenk, C.; Jenkins, A.
  2006, \mnras, 367, 1039

\bibitem[]{Hop06}
Hopkins P.F., Hernquist L., Cox T.J., Di Matteo T., Robertson B.,
Springel V., 2006, \apjs, 163, 1

\bibitem[]{Hop07}
Hopkins P.F., Hernquist L., Cox T.J., Keres D., 2007, preprint
(arXiv:0706.1243)

\bibitem[]{jmb98}
Jing, Y. P., Mo, H. J., \& B{\" o}rner, G.\ 1998, \apj, 494, 1

\bibitem[]{Jing07}
Jing, Y. P., Suto, Y., Mo, H. J., 2007 , \apj, 657, 664

\bibitem[]{}
Keselman J.A., Nusser A., 2007, \mnras, preprint (arXiv:0707.4361)

\bibitem[]{Lac93}
Lacey C., Cole S., 1993, \mnras, 262, 627


\bibitem[]{Li06}
Li C., Kauffmann G., Jing Y.P., White S.D.M., B$\ddot{o}$rner G.,
Cheng F.Z., 2006, \mnras, 368 , 21

\bibitem[]{Li07}
Li C., Jing Y.P., Kauffmann G., B$\ddot{o}$rner G., Kang X.,
Wang L., 2007a, \mnras, 376 ,984

\bibitem[]{LiY07}
Li Y., Mo H.J., van den Bosch F.C., Lin W.P., 2007b, \mnras, 379, 689

\bibitem[]{Mag03}
Magliocchetti M., Porciani C., 2003, \mnras , 346, 186

\bibitem[]{Mo92}
Mo H. J., Jing Y.P., B\"orner G., 1992, \apj, 392, 452

\bibitem[]{MoW96}
Mo H. J., \& White S.D.M. 1996, \mnras , 282, 347

\bibitem[]{Nag04}
Nagashima, M., Yoshii, Y., 2004, \apj , 610, 23

\bibitem[]{Nei06}
Neistein E., van den Bosch F.C., Dekel A., 2006, \mnras, 372, 933

\bibitem[]{Pad04}
Padilla N.D., et al., 2004, \mnras , 352 ,211

\bibitem[]{2000MNRAS.318.1144P}
Peacock J.~A., Smith R.~E., 2000, MNRAS, 318, 1144

\bibitem[]{PS74}
Press W.H., Schechter P., 1974, \apj , 187, 425

\bibitem[]{Ree07}
Reed D.S., Governato F., Quinn T., Stadel J., Lake G., 2007, \mnras , 378, 777

\bibitem[]{saunders00a}
Saunders W. {et~al.} 2000, \mnras, 317, 55

\bibitem[]{2000MNRAS.318..203S}
Seljak U., 2000, MNRAS, 318, 203

\bibitem[]{Sel04}
Seljak U., Warren M.S., 2004, \mnras , 355, 129

\bibitem[]{ST99}
Sheth, R.K., Tormen, G., 1999, \mnras , 308, 119

\bibitem[]{SMT01}
Sheth, R.K., Mo H.J., Tormen, G., 2001, \mnras , 323, 1

\bibitem[]{ST04}
Sheth, R.K., Tormen, G., 2004, \mnras, 350, 1385

\bibitem[]{Sp05}
Springel, V., et al., 2005, Nature, 435, 629

\bibitem[]{Tin05}
Tinker J.L., Weinberg D.H., Zheng Z., Zehavi I., 2005, \apj, 631, 41

\bibitem[]{Tin07}
Tinker, J. L., Conroy, C., Norberg, P., Patiri, S. G.,
  Weinberg, D.  H., Warren, M. S., 2007, arXiv0707.3445

\bibitem[]{vale06}
Vale A., Ostriker J.P., 2006, \mnras , 371, 1173

\bibitem[]{vdB02}
van den Bosch F.C., 2002, \mnras, 331, 98

\bibitem[]{BYM03}
van den Bosch F.C., Yang X., Mo H.J., 2003, \mnras , 340, 771

\bibitem[]{B07}
van den Bosch F.C., Yang X., Mo H.J., Weinmann S.M., Maccio A.,
More S., Cacciato M., Skibba R., Kang X., 2007, \mnras , 376, 841

\bibitem[]{WMJ07}
Wang  H.Y., Mo  H.J., Jing  Y.P., 2007, \mnras, 375, 633

\bibitem[]{Wec02}
Wechsler R.H., Bullock J.S., Primack J.R., Kravtsov, A. V.,
Dekel A., 2002 , \apj, 568, 52

\bibitem[]{Wec06}
Wechsler R.H., Zentner A.R., Bullock J.S., Kravtsov A.V.,
Allgood B., 2006 , \apj, 652 , 71

\bibitem[]{WBYM06}
Weinmann S.M., van den Bosch F.C., Yang X., Mo H.J., 2006a, \mnras, 366, 2

\bibitem[]{Wei06}
Weinmann S.M., van den Bosch F.C., Yang X., Mo H.J., Croton D.J.,
Moore B., 2006b, \mnras, 372, 1161

\bibitem[]{Wet07}
Wetzel A.R., Cohn J.D., White M., Holz, D.E.,  Warren M.S., 2007,
\apj, 656, 139

\bibitem[]{YMB03}
Yang, X., Mo, H. J., \& van den Bosch, F. C. 2003, \mnras , 339, 1057

\bibitem[]{YMB04}
Yang, X., Mo, H. J., Jing, Y. P., van den Bosch, F. C., Chu, Y. Q. ,
2004 ,\mnras , 350 , 1153

\bibitem[]{Y05a}
Yang, X., Mo, H. J., Jing, Y. P., van den Bosch, F. C., Jing, Y. P.,
2005a, \mnras , 356, 1293

\bibitem[]{Y05b}
Yang, X., Mo, H. J., Jing, Y. P., \& van den Bosch, F. C., Jing, Y.
P., 2005b, \mnras , 357, 608

\bibitem[]{YMB06}
Yang, X., Mo, H. J., \& van den Bosch, F. C. 2006, ApJL, 638, 55

\bibitem[]{}
Yang, X., Mo, H. J., van den Bosch, F. C., Pasquali, A., Li, C.,
  Barden, M., 2007, ApJ, 671, 153

\bibitem[]{york00}
York, D.~G., et al., 2000, \aj, 120, 1579

\bibitem[]{Zan03}
Zandivarez, A., Merch$\acute{a}$n, M. E., Padilla, N. D.,
  2003, \mnras , 344 ,247

\bibitem[]{Zhe05}
Zheng Z., et al., 2005, \apj , 633, 791

\end{thebibliography}
\end{document}